\documentclass[twocolumn, graphics, floatfix, a4paper, aps, prx, superscriptaddress, longbibliography, showpacs]{revtex4-1}
\usepackage{graphicx}
\usepackage{bm}
\usepackage[usenames,dvipsnames]{xcolor}
\usepackage[colorlinks=true,urlcolor=blue,citecolor=blue,linkcolor=blue]{hyperref}
\usepackage{amssymb,ulem,amsmath}
\usepackage{longtable}
\newcommand{\bra}[1]{\left\langle #1\right|}
\newcommand{\ket}[1]{\left|#1\right\rangle}

\newcommand{\qql}{\textquotedblleft}
\newcommand{\qqr}{\textquotedblright}
\newcommand{\vc}[1]{\bm{\mathrm{#1}}}

\begin{document}
\title{Preformed pairs in flat Bloch bands}
\author{Murad Tovmasyan}
\affiliation{Institute for Theoretical Physics, ETH Zurich, 8093 Z\"urich, Switzerland}

\author{Sebastiano Peotta} 
\affiliation{COMP Centre of Excellence, Department of Applied Physics, Aalto University School of Science, FI-00076 Aalto, Finland}

\author{Long Liang}
\affiliation{COMP Centre of Excellence, Department of Applied Physics, Aalto University School of Science, FI-00076 Aalto, Finland}

\author{P\"aivi T\"orm\"a}
\affiliation{COMP Centre of Excellence, Department of Applied Physics, Aalto University School of Science, FI-00076 Aalto, Finland}

\author{Sebastian D. Huber} 
\affiliation{Institute for Theoretical Physics, ETH Zurich, 8093 Z\"urich, Switzerland}

\begin{abstract}
In a flat Bloch band the kinetic energy is quenched and single particles cannot propagate since they are localized due to destructive interference. Whether this remains true in the presence of interactions is a challenging question because a flat dispersion usually leads to highly correlated ground states. Here we compute numerically the ground state energy of lattice models with completely flat band structure in a ring geometry in the presence of an attractive Hubbard interaction. We find that the energy as a function of the magnetic flux threading the ring has a half-flux quantum $\Phi_0/2 = hc/(2e)$ period, indicating that only bound pairs of particles with charge $2e$ are propagating, while single quasiparticles with charge $e$ remain localized. For some one dimensional lattice models we show analytically that in fact the whole many-body spectrum has the same periodicity. Our analytical arguments are valid for both bosons and fermions, for generic interactions respecting some symmetries of the lattice and at arbitrary temperatures. Moreover for the same one dimensional lattice models we construct an extensive number of exact conserved quantities. These conserved quantities are associated to the occupation of localized single quasiparticle states and force the single-particle propagator to vanish beyond a finite range. Our results suggest that in lattice models with flat bands preformed pairs dominate transport even above the critical temperature of the transition to a superfluid state.
\end{abstract}
\maketitle
\section{Introduction}

A flat band is a Bloch band with constant energy as a function of quasimomentum and, as a consequence, it is highly degenerate. In a flat band it is always possible to find a basis composed of eigenstates that are completely localized, a possible choice are the Wannier functions~\cite{Marzari:2012}. Only in a flat band the Wannier functions are eigenstates of the single-particle Hamiltonian. A particle initially placed in one of these states will remain localized for arbitrary long times~\cite{Vidal:1998}. In other words, a flat band is insulating even if partially filled, in contrast to a dispersive band. 

Interactions can have the effect of enabling transport, for example an attractive interaction of the Hubbard type can induce a nonzero superfluid weight in a flat band which is proportional to the interaction strength~\cite{Peotta:2015,Julku:2016,Tovmasyan:2016}. 
As shown in Ref.~\cite{Vidal:2000} and also in a recent work by the some of the authors~\cite{Tovmasyan:2016}, interactions can lead to the formation of two-body bound states with finite effective mass out of two bare particles, which have infinite effective mass due to the band flatness. The finite effective mass of the bound pairs is proportional to the interaction strength and to the overlap between the Wannier functions of the flat band. This overlap is measured by a band structure invariant, the quantum metric~\cite{Tovmasyan:2016}. The formation of such mobile pairs is the mechanism at the root of the finite superfluid weight in a flat band.

\begin{figure}
	\includegraphics{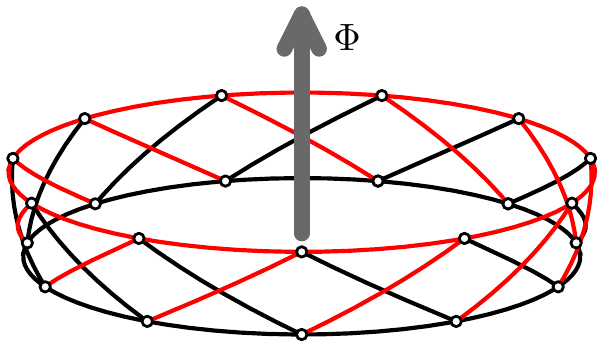}
	\caption{\label{fig:ring_geometry} Illustration of the ring geometry considered in this work using the example of the Creutz ladder introduced below, see Fig.~\ref{fig:lattice_models}\textbf{a}. The twist in the boundary condition along the non-contractible loop is controlled by the magnetic flux $\Phi$ threading the ring. Even if the magnetic field is strictly zero in the region of the lattice, the magnetic flux can affect the physical properties in the ring, in particular induce a persistent ground state current, a phenomenon known as the Aharonov-Bohm effect. In this work we are interested in the periodicity of the ground state energy $E_0(\Phi)$ as a function of the magnetic flux, as a tool to probe the nature of the charge carriers of the persistent current, in particular if they are single particles, or composite objects (Cooper pairs).}
\end{figure}

Interest in flat band superfluidity has arisen from the prediction, based on Bardeen-Cooper-Schrieffer (BCS) theory, that the critical temperature is dramatically enhanced in a flat band~\cite{Kopnin:2011,Heikkila:2011}. However it is unclear to what extent BCS theory can be applied in the case of flat bands which are often not amenable to a mean-field treatment (see however Ref.~\cite{Tovmasyan:2016}). A crucial step towards a better estimate of the critical temperature is a better understanding of the nature of the excitations above the ground state in a flat band superfluid. In the case of a flat band superfluid the normal state above the critical temperature is expected not to be a Landau-Fermi liquid, which is characterized by long-lived fermionic quasiparticle excitations. Indeed by adiabatically switching off the interparticle interaction the system becomes an insulator rather than a noninteracting Fermi gas with a well defined Fermi surface, and this precludes the usual Landau-Fermi liquid scenario~\cite{Vignale_book}. In this work we substantiate this expectation by showing rigorously that in selected lattice models with flat bands the only charge carriers are bosonic two-body bound states, which are responsible for any kind of transport (dissipative and superfluid), while fermionic single quasiparticles remain localized as in the absence of interactions.
Morever we show both numerically and analytically that the composite nature of the charge carriers manifests in the Aharonov-Bohm effect.

The Aharonov-Bohm effect~\cite{Ehrenberg:1949,Aharonov:1959,Aharonov:1961} is an exquisite quantum mechanical phenomenon where a quantum particle propagating coherently along a ring is sensitive to the magnetic flux threading the ring even if the magnetic field and thus the classical force acting on the particle is locally zero. An example of such a ring geometry for one of the lattice models considered in this work is shown in Fig.~\ref{fig:ring_geometry}. In particular for nonzero magnetic flux time-reversal symmetry is broken and the ground state of the particle possesses a nonzero persistent current. The current is a periodic function of the magnetic flux with period given by the flux quantum $\Phi_0 = hc/e$ (in Gaussian units), where $e$ is the electron charge. Persistent currents have been observed in mesoscopic metallic rings~\cite{Webb:1985,Washburn:1986,Bachtold:1999,Grbic:2008,Bleszynski-Jayich:2009}; see Ref.~\cite{Viefers:2004} for a review on the subject. The periodicity of the persistent current as a function of the magnetic flux provides important information regarding the nature of the charge carriers. Indeed the observation that in superconducting loops the flux is quantized in units of $\Phi_0/2 = hc/(2e)$ (the superconducting flux quantum) is a clear-cut confirmation of BCS theory, which regards the condensation of bound pairs of electrons (Cooper pairs) as the origin of superconductivity~\cite{Deaver:1961,Doll:1961}.
In frustrated Josephson junction arrays a supercurrent with period $\Phi_0/4$ has been measured, indicating that pairs of Cooper pairs (bound states with charge $4e$) are responsible for transport~\cite{Pop:2008,Gladchenko:2008}.


Here we use the Aharonov-Bohm effect as a tool for probing the nature of the charge carriers in lattices with flat bands.
To this end we compute numerically, using density matrix renormalization group (DMRG) and exact diagonalization (ED), the ground state energy as a function of the magnetic flux $E_{0}(\Phi)$ in a ring geometry as in Fig.~\ref{fig:ring_geometry}. We consider models whose entire band structure is composed of flat bands, the dice lattice analyzed by Vidal et al.~\cite{Vidal:1998}, the one dimensional Creutz ladder~\cite{Creutz:1999,Tovmasyan:2013,Tovmasyan:2016,Takayoshi:2013} and the diamond chain~\cite{Vidal:2000,Doucot:2002}. In the numerical calculations we add the attractive Hubbard interaction to the noninteracting lattice Hamiltonian. In the case of the attractive Hubbard interaction it has been established that the ground state is superfluid if the partially filled flat band has nonzero quantum metric~\cite{Peotta:2015,Julku:2016,Tovmasyan:2016,Liang:2017a,Liang:2017b}. All the bands of the lattice models considered in this work satisfy this condition. This implies that  $[E_0(\Phi)-E_0(0)]\cdot L^{2-d}$ scales to a non-constant function in the thermodynamic limit $L\to +\infty$ ($L$ is the linear size of the system and $d = 1,2,3$ the dimensionality). We often find that the ground state energy as a function of flux has a half-flux quantum ($\Phi_0/2$) periodicity, namely $E_0(\Phi) = E_0(\Phi+ \Phi_0/2)$ within numerical accuracy, indicating that the charge carriers are bound states of charge $2e$. We show explicitly that when the bands are not flat this is not true anymore in general and the energy is simply $\Phi_0$-periodic as required by gauge invariance, but not $\Phi_0/2$-periodic.

The numerical results are confirmed by analytical arguments for some one dimensional lattice Hamiltonians. Specifically, we construct a unitary transformation $\mathcal{\hat U}$ that intertwines the many-body Hamiltonians whose magnetic fluxes differ by $\Phi_0/2$, namely 
\begin{equation}\label{eq:intertwining_op}
\mathcal{\hat U}\mathcal{\hat H}(\Phi)\mathcal{\hat U}^\dagger = \mathcal{\hat H}(\Phi+\Phi_0/2)\,.
\end{equation}
Since the operators $\mathcal{\hat H}(\Phi)$ and $\mathcal{\hat H}(\Phi+\Phi_0/2)$ are related by a similarity transformation $\mathcal{\hat U}$ they have the same spectrum, in particular the ground state energy is the same $E_{0}(\Phi) = E_{0}(\Phi+\Phi_0/2)$. This analytical argument is very powerful since it is valid for both bosons and fermions and for general interaction terms that preserve certain symmetries of the lattice, as explained in the following. It is independent of the sign of the interaction and shows that the entire spectrum of the Hamiltonian has the same periodicity, implying that the $\Phi_0/2$-periodicity should be present in all observable quantities also at finite temperatures, the most important being the persistent current.

We find that the unitary transformation $\mathcal{\hat U}$ is local and its translation by an arbitrary number of unit cells is an equivalent but distinct transformation $\mathcal{\hat U}'$. Therefore their product $\mathcal{\hat U}\mathcal{\hat U}'$ is a nontrivial symmetry of the Hamiltonian since $\mathcal{\hat H}(\Phi)$ and $\mathcal{\hat H}(\Phi+\Phi_0)$ are physically equivalent. In fact there is an extensive number of such symmetry operators. Physically these conserved quantities are associated to the occupation of Wannier states by unpaired particles and imply that the single-particle propagator is vanishing beyond a finite range. Thus, at least in one dimension, we can conclude that unpaired particles in flat bands remain localized even in the presence of interactions. We recover in this alternative way the local $\mathbb{Z}_2$ symmetries of the Bose-Hubbard model on the diamond chain, first found in Ref.~\cite{Doucot:2002}. For all the other lattices considered here, the existence of an extensive number of local symmetries is a new result.

Our analytical arguments in one dimension cannot be extended straightforwardly to higher dimensions, for example to the 2D dice lattice. However, for this lattice we find numerically almost the same behavior as in one dimension. This is a hint that charge transport in systems with flat bands is generally dominated by preformed pairs even above the critical temperature of the transition to a superfluid state. Our results represent an essential starting point on the way to reach more general conclusions in this sense.
Moreover, our rigorous analytical arguments for specific models can be used as a benchmark for approximate methods designed to capture the properties of systems with such preformed pairs, in particular unconventional superconductors~\cite{Randeria:1998,Rice:2012,Stewart:2017}. 
As it is discussed in Sec.~\ref{sec:disc_concl}, our results are experimentally relevant also for ultracold gases in optical lattices and artificially engineered lattices in the solid state context.


The work is organized as follows. In Sec.~\ref{sec:lattice_Hamiltonians} the specific lattice Hamiltonians studied in this work are introduced along with some essential notation. In Sec.~\ref{sec:boundary_condition} we review the concept of twisted boundary condition in some detail, in particular in the case of lattice Hamiltonians with complex orbital structure (multiorbital/multiband lattice Hamiltonians). In Sec.~\ref{sec:numerics} the original numerical results are presented. We focus on the ground state energy $E_0(\Phi)$ as a function of the magnetic flux through the ring (see Fig.~\ref{fig:ring_geometry}) and study its periodicity in interacting lattice Hamiltonians. In Sec.~\ref{sec:graph_automorphism} we introduce the intertwining operators $\mathcal{\hat U}$ which rigorously prove that the spectrum of an interacting Hamiltonian is $\Phi_0/2$-periodic according to Eq.~(\ref{eq:intertwining_op}) and explain how they are associated to local graph automorphism of the lattice. This method can be applied to the Creutz ladder, the diamond chain and to the one-dimensional reduction of the dice lattice for some special values of the hopping matrix elements. In Sec.~\ref{sec:conserved quantities} the intertwining operators are used to construct an extensive number of local and mutually commuting conserved quantities. In Sec.~\ref{sec:interpretation} we interpret these conserved quantities as the parity of the occupation number of Wannier functions with support on a finite number of lattice sites. In Sec.~\ref{sec:projected_conserved_quantities} we show that the operators encoding the conserved quantities commute also with the Hamiltonian projected on the many-body subspace of a chosen flat band, and using this fact we explain some features of the effective low energy theory of the Creutz ladder first pointed out in our previous work~\cite{Tovmasyan:2016}. The detailed description of the lattice models is provided in the appendices, along with some useful results relevant to the main text. 

In the following we set $\hbar = e = c = 1$, which means that the flux quantum is $\Phi_0 = 2\pi$ in our units.

\section{Lattice Hamiltonians}
\label{sec:lattice_Hamiltonians}

We consider several lattice models with a common feature: the band structure of the noninteracting (quadratic) Hamiltonian is composed only of perfectly flat bands. The noninteracting part of the various many-body Hamiltonians takes the generic form
\begin{equation}\label{eq:free_H}
\mathcal{\hat H}_{0} = \sum_{\vc{i},\vc{j}}\hat{\vc{c}}^\dagger_{\vc{i}} K(\vc{i}-\vc{j}) \hat{\vc{c}}_{\vc{j}} = \sum_{\vc{k}}\hat{\vc{c}}^\dagger_{\vc{k}} \widetilde{K}(\vc{k}) \hat{\vc{c}}_{\vc{k}}\,.
\end{equation}
Since all the lattices considered here have more than one orbital per unit cell ($N_{\rm orb} > 1$), we denote by $\hat{\vc{c}}_{\vc{i}} = (\hat{c}_{\vc{i},1},\hat{c}_{\vc{i},2},\dots,\hat{c}_{\vc{i},\alpha = N_{\rm orb}})^T$ the vector of fermionic or bosonic annihilation operators relative to all the orbitals in unit cell $\vc{i}$. The orbitals within a unit cell are labelled by Greek letters $\alpha,\beta = 1,2,\dots,N_{\rm orb}$, while multi-indices $\vc{i} = (i_1,i_2,\dots,i_d)\in \mathbb{Z}^{d}$ are used to label the unit cells of the lattice. We consider only one dimensional $d = 1$ and two dimensional $d = 2$ lattices in this work. To each unit cell is associated a unique lattice vector $\vc{r}_{\vc{i}} = i_1\vc{a}_1+i_{2}\vc{a}_2+\ldots +i_d\vc{a}_d$ of the Bravais lattice. Here $\vc{a}_{j = 1,2,\ldots, d}$ are the fundamental lattice vectors of the Bravais lattice. The hopping matrix elements of the lattice are encoded in the matrix $K(\vc{i}-\vc{j}) = [K(\vc{j}-\vc{i})]^\dagger$ (a $N_{\rm orb}\times N_{\rm orb}$ matrix for fixed argument), which depends only on the difference $\vc{i}-\vc{j}$, reflecting the discrete translational invariance of the lattice. It is useful to present the noninteracting Hamiltonian using the hopping matrix in momentum space $\widetilde{K}(\vc{k}) = [\widetilde{K}(\vc{k})]^\dagger = \sum_{\vc{j}} e^{-i\vc{k}\cdot \vc{r}_{\vc{j}}}K(\vc{j})$, as on the right-hand side of Eq.~(\ref{eq:free_H}). The field operators in momentum space $\hat{\vc{c}}_{\vc{k}} = (\hat{c}_{\vc{k},1},\hat{c}_{\vc{k},2},\dots,\hat{c}_{\vc{k},\alpha = N_{\rm orb}})^T$ are defined by the expansion $\hat{\vc{c}}_{\vc{i}} = \frac{1}{\sqrt{N}_{\rm c}}\sum_{\vc{k}}e^{i\vc{k}\cdot\vc{r}_{\vc{i}}}\hat{\vc{c}}_{\vc{k}}$. Here $N_{\rm c}$ is the total number of unit cells in a finite lattice. The total number of lattice sites is then $N_{\rm sites} = N_{\rm c}N_{\rm orb}$. The $\vc{k}$-dependent $N_{\rm orb}\times N_{\rm orb}$ matrices $\widetilde{K}(\vc{k})$ for all the lattice models considered in this work are provided in Appendix~\ref{app:lattice_Ham}.

\begin{figure}
	\centering\includegraphics{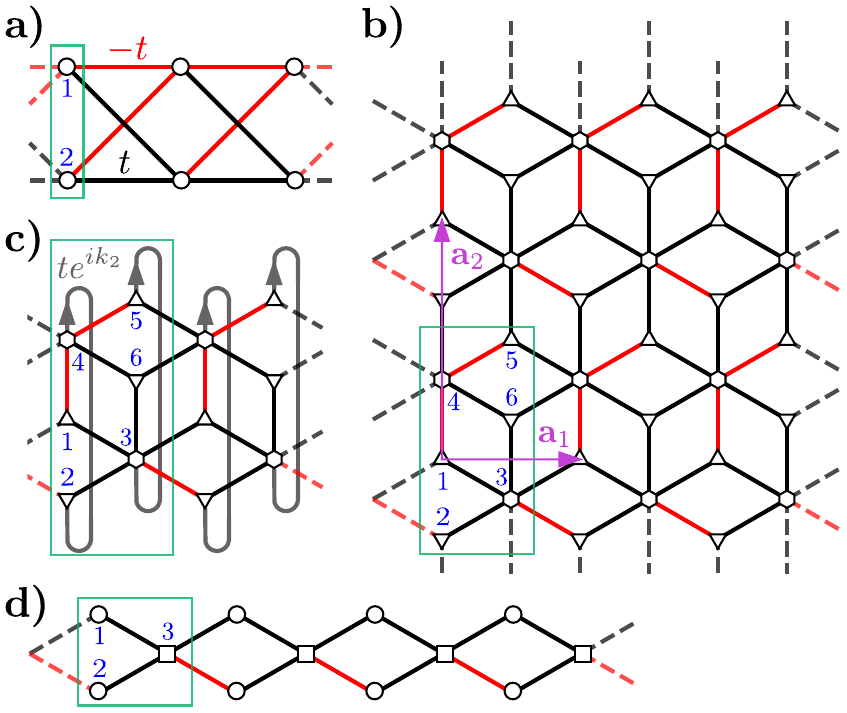}
	\caption{Lattice structure and labelling conventions for the (\textbf{a}) Creutz ladder, 2D (\textbf{b}) and 1D (\textbf{c}) dice lattices and diamond chain (\textbf{d}). The green rectangles are the (magnetic) unit cell. The black lines correspond to hopping matrix elements with amplitude $t>0$, while the red lines have opposite amplitude. In the 1D dice lattice (\textbf{c}) the sense of the complex hopping matrix elements (grey lines) is indicated by an arrow.\label{fig:lattice_models} }
\end{figure}

The simplest model studied here is the Creutz ladder, which we have considered in our previous works~\cite{Tovmasyan:2013,Tovmasyan:2016}. The Creutz ladder is shown in Fig.~\ref{fig:lattice_models}\textbf{a}. It is the simplest lattice with a completely flat spectrum, in fact a lattice with $N_{\rm orb} = 1$ and whose single band is perfectly flat corresponds to the case $K(\vc{i}-\vc{j}) = 0$, a trivial lattice of uncoupled sites. The Creutz ladder realizes the first nontrivial case ($N_{\rm orb}=2$) of a lattice model with a completely flat band structure. 

We use here a representation of the Creutz ladder where time-reversal symmetry is manifest, that is all the hopping matrix elements [the matrix elements of $K(\vc{i}-\vc{j})$] are real (see Fig.~\ref{fig:lattice_models}). This representation is related to the one used in other works~\cite{Creutz:1999,Takayoshi:2013,Tovmasyan:2013,Tovmasyan:2016} by a gauge transformation. In this work we call a gauge tranformation a canonical transformation of the field operators of the form
\begin{equation}\label{eq:gauge}
\hat{c}_{\vc{i}\alpha} \to \hat{c}_{\vc{i}\alpha}e^{i\phi(\vc{i},\alpha)}\,.
\end{equation}
The phase $\phi(\vc{i},\alpha)$ is an arbitrary function of the unit cell index $\vc{i}$ and of the orbital index $\alpha$. This transformation changes the hopping matrix elements $K(\vc{i}-\vc{j})$ but does not affect any physical property.

The 2D dice lattice~\cite{Vidal:1998,Moeller:2012} 
is shown in Fig.~\ref{fig:lattice_models}\textbf{b}. The modulation of the phase of the hopping matrix elements, as shown in Fig.~\ref{fig:lattice_models}\textbf{b}, corresponds to a uniform perpendicular magnetic field with flux commensurate to the lattice, equal to half-flux quantum for each elementary rhombus. Our convention for the magnetic unit cell is also shown in Fig.~\ref{fig:lattice_models}\textbf{b}. By performing the Fourier transform of $\widetilde{K}(\vc{k})$ only with respect to $\vc{k}\cdot \vc{a}_1 = k_1$ (see Eq.~(\ref{eq:partial_FT}) in Appendix~\ref{app:2D_dice_Ham}), one obtains a family of one dimensional lattices parametrized by a continuous parameter $k_2 \in (-\pi,\pi]$, shown in Fig.~\ref{fig:lattice_models}\textbf{c}. We call these the one dimensional reductions of the 2D dice lattice, or 1D dice lattices for brevity.

By eliminating sites $\alpha = 4,5,6$ of the 2D dice lattice one is left with decoupled 1D lattices all identical to the diamond chain of Ref.~\cite{Doucot:2002}, shown in Fig.~\ref{fig:lattice_models}\textbf{d}. A more general version of the 2D dice lattice with completely flat spectrum is presented in Appendix~\ref{app:2D_dice_Ham}. Using this general model it is possible to continuously interpolate between the 2D dice lattice and the diamond chain.

All of these noninteracting lattice Hamiltonians have a band structure composed only of completely flat bands. In Appendix~\ref{app:compact_Wannier} we provide particularly convenient bases for the degenerate subspaces of the various flat bands. These bases are composed of Wannier functions which are localized only on a finite number of unit cells, that is they are compactly localized. These compactly localized Wannier functions are important for our purposes since they are associated to the conserved quantities to be presented in Sec~\ref{sec:conserved quantities}.

\begin{figure}
\includegraphics[scale = 0.6]{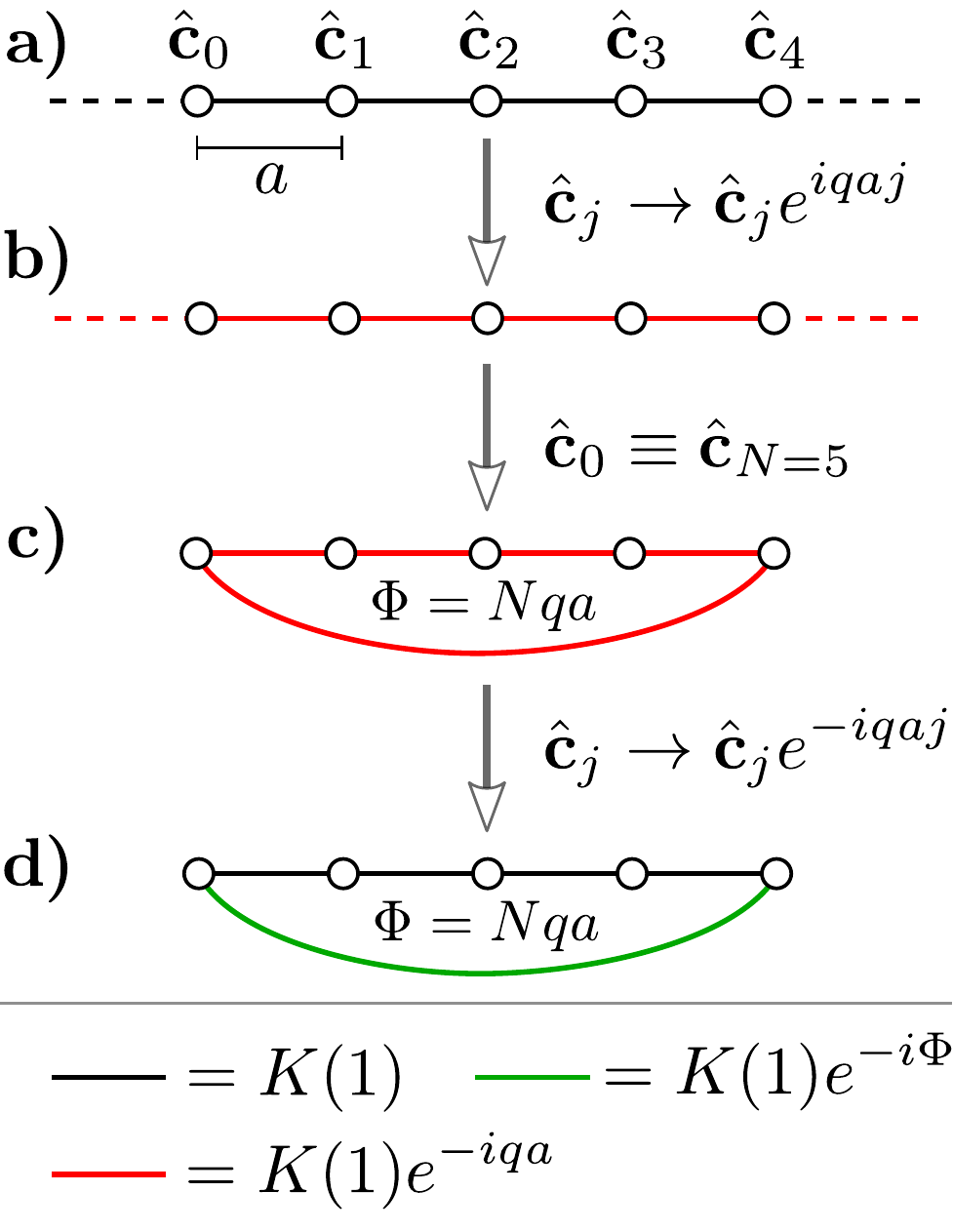}
\caption{\label{fig:flux_insertion} Illustration of flux insertion by folding in a one dimensional system. Each circle represents a unit cell and $\hat{\bm{\mathrm{c}}}_j = (\hat{c}_{j,\alpha = 1},\dots,\hat{c}_{j,\alpha = N_{\rm orb}})^T$ is the associated vector of field operators. The infinite lattice in \textbf{a)} represents a quadratic Hamiltonian defined by hopping matrix elements $K(j)$ (\ref{eq:free_H}). The black lines represent the nearest neighbour hopping matrix elements $K(1)$ (and their Hermitian conjugates $K(-1)$) as indicated by the color code at the bottom of the figure. \textbf{b)} After the gauge transformation (\ref{eq:gauge_flux_insertion}) the hopping terms acquire a phase (\ref{eq:flux_insertion}). The modified hopping terms are indicated by red lines. The gauge transformation leaves the Hamiltonian in a translational invariant form and does not affect the system properties. \textbf{c)} By identifying unit cells (\ref{eq:unit_cell_ident}) one obtains a finite system of size $N$. The flux $\Phi = Nqa$ through the ring does affect the physical properties, for example the ground state current along the ring. \textbf{d)} By performing the inverse of the gauge transformation in step \textbf{a-b} one obtains an equivalent Hamiltonian where only the hopping terms connecting unit cells $j = 0$ and $j = N-1$ acquire a phase (represented by the green line). The presence of the magnetic flux $\Phi$ is equivalent to twisted boundary conditions in this case.}
\end{figure}

\section{Twisted boundary conditions}
\label{sec:boundary_condition}

In this work we focus on the change of the ground state energy $E_0(\Phi)$ with the magnetic flux $\Phi$ in a ring geometry, as in Fig.~\ref{fig:ring_geometry}. The magnetic flux is equivalent to a twist in the boundary conditions. In the following it is explained how twisted boundary conditions are introduced in a general multiorbital lattice Hamiltonian. The case of a two dimensional lattice is considered for definiteness. 
The same procedure is illustrated in Fig.~\ref{fig:flux_insertion} for a one dimensional chain.
The starting point is a finite-size lattice with Born-von K\'arm\'an periodic boundary conditions. This is obtained from Eq.~(\ref{eq:free_H}) by identifying the unit cell labels $\vc{i}$, and the field operators $\hat{\vc{c}}_{\vc{i}}$, that differ by integer multiples of $N_1\vc{\hat e}_1$ and $N_2\vc{\hat e}_2$, where $\vc{\hat e}_1 = (1,0)^T$ and $\vc{\hat e}_2 = (0,1)^T$ are unit vectors and the positive integers $N_1,\,N_2$ fix the linear sizes of the system along the two independent directions of the Bravais lattice. In other words
\begin{equation}\label{eq:unit_cell_ident}
\hat{\vc{c}}_{\vc{i}} \equiv \hat{\vc{c}}_{\vc{j}} \quad \text{if} \quad \vc{i}-\vc{j} = m_1N_1\vc{\hat{e}}_1 + m_2N_2\vc{\hat{e}}_2 
\end{equation}
with $m_1,m_2 \in \mathbb{Z}$. 
Then the Hamiltonian of the finite system reads
$\mathcal{\hat H}_{0} = \sum_{(\vc{i},\vc{j})}\hat{\vc{c}}^\dagger_{\vc{i}} K_{\rm finite}(\vc{i},\vc{j}) \hat{\vc{c}}_{\vc{j}}$
where the sum $\sum_{(\vc{i},\vc{j})}$ runs over each inequivalent pair only once. The hopping matrix $K_{\rm finite}(\vc{i},\vc{j})$ in the finite system is equal to the hopping matrix in the infinite system $K(\vc{i}-\vc{j} + m_1N_1\vc{\hat e}_1 + m_2N_2\vc{\hat e}_2)$ for the only pair of values $m_1,m_2 \in \mathbb{Z}$ for which it is nonzero. This assumes that the range of $K(\vc{i}-\vc{j})$ is less than the size of the system as in the case of the Hamiltonians considered in this work since only $K(\vc{0}),\,K(\pm\vc{\hat e}_1),\,K(\pm\vc{\hat e}_2)$ are nonzero.

We call this procedure folding and the result
is a finite-size lattice with the topology of a torus in 2D, or ring in 1D, see Fig.~\ref{fig:flux_insertion}\textbf{b-c}. It is possible to introduce magnetic fluxes threading the two noncontractible loops of the torus by \textit{first} modifying the hopping matrix elements in the infinite lattice as follows (see Fig.~\ref{fig:flux_insertion}\textbf{a-b})
\begin{equation}\label{eq:flux_insertion}
K(\vc{i}-\vc{j}) \to K(\vc{i}-\vc{j})e^{-i\vc{q}\cdot(\vc{r}_{\vc{i}}-\vc{r}_{\vc{j}})}
\end{equation}
which corresponds to a gauge transformation 
\begin{equation}\label{eq:gauge_flux_insertion}
\hat{c}_{\vc{i}\alpha} \to \hat{c}_{\vc{i}\alpha}e^{i\vc{q}\cdot\vc{r}_{\vc{i}}}\,,
\end{equation}
and \textit{then} folding the infinite lattice as explained above.
Even if Eq.~(\ref{eq:gauge_flux_insertion}) is a gauge transformation in the infinite system, after the folding one obtains a family of physically distinct finite-size systems (Fig.~\ref{fig:flux_insertion}\textbf{c}).
%
%
Indeed the magnetic flux threading the non-contractible loops along the $\vc{a}_1$ direction is $\Phi_1 = N_1\vc{a}_1\cdot \vc{q}$. Similarly, for the non-contractible loops along direction $\vc{a}_2$ the magnetic flux is $\Phi_2 = N_2\vc{a}_2\cdot\vc{q} $. On the other hand the magnetic flux through all contractible loops is unaffected by the transformation~(\ref{eq:flux_insertion}). The family of physically distinct finite-size 2D Hamiltonians $\mathcal{\hat H}_0(\Phi_1,\Phi_2)$ is parameterized by the magnetic fluxes $\Phi_1,\,\Phi_2 \in (-\pi,\pi]$. In a lattice the magnetic flux is defined modulo $\Phi_0 = 2\pi$ in our units, since the shift $\Phi \to \Phi+2\pi$ can always be implemented by a gauge transformation. 

The flux-insertion transformation obtained from Eq.~(\ref{eq:gauge_flux_insertion}) and the subsequent folding of the infinite lattice has the advantage of explicitly preserving the translational symmetry of the lattice Hamiltonian. By performing a gauge transformation which breaks translational invariance in the finite system (after folding) it is possible to localize the flux-insertion transformation, that is only the hoppings that cross a specified boundary are modified with respect to the zero flux case. A possible gauge transformation of this kind is simply the inverse of Eq.~(\ref{eq:gauge_flux_insertion}), namely $\hat{c}_{\vc{i}\alpha} \to \hat{c}_{\vc{i}\alpha}e^{-i\vc{q}\cdot\vc{r}_{\vc{i}}}$ (see Fig.~\ref{fig:flux_insertion}\textbf{c-d}). As a result the hopping terms that cross the boundary between unit cells labelled by $(N_1-1,j)$ and those labelled by $(N_1,j) \equiv (0,j)$ in the finite system are modified in the local flux-insertion transformation, for example
\begin{equation}\label{eq:flux_insertion_boundary_1}
t\hat{c}^\dagger_{\vc{0},\alpha} \hat{c}_{(N_1-1)\vc{\hat e}_1,\beta} \to 
t\hat{c}^\dagger_{\vc{0},\alpha} \hat{c}_{(N_1-1)\vc{\hat e}_1,\beta}e^{-i\Phi_1}\,.
\end{equation}
The same occurs to the terms crossing the boundary between unit cells $ (j,N_2)\equiv(j,0)$ and $(j,N_2-1)$, for example
\begin{equation}\label{eq:flux_insertion_boundary_2}
t\hat{c}^\dagger_{\vc{0},\alpha} \hat{c}_{(N_2-1)\vc{\hat e}_2,\beta} \to 
t\hat{c}^\dagger_{\vc{0},\alpha} \hat{c}_{(N_2-1)\vc{\hat e}_2,\beta}e^{-i\Phi_2}\,.
\end{equation}
All the hopping terms that do not cross the two boundaries are unaffected by the local flux-insertion transformation. Therefore Eqs.~(\ref{eq:flux_insertion_boundary_1})-(\ref{eq:flux_insertion_boundary_2}) correspond to a twist of the boundary conditions, as shown in Fig.~\ref{fig:flux_insertion}\textbf{d}. It is the local form of the flux-insertion transformation that will be most useful in the following. From the point of view of physical properties the two possibilities, shown in Fig.~\ref{fig:flux_insertion}\textbf{c} and \ref{fig:flux_insertion}\textbf{d}, are completely equivalent. 

\begin{figure}
	\includegraphics{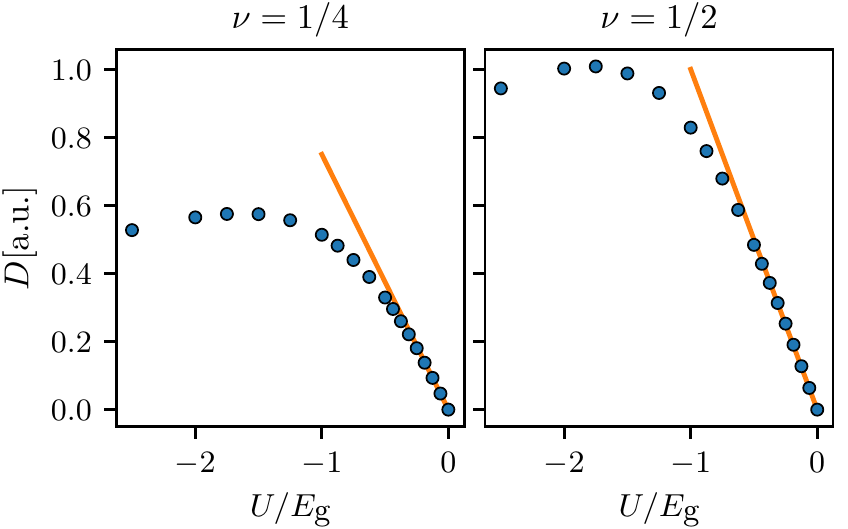}
	\caption{\label{fig:Drude_Creutz}Drude weight of the Creutz ladder from DMRG and comparison to the exact result of Eq.~(\ref{eq:exact_Ds}) in the isolated flat band limit $|U| \ll E_{\rm g}$ (straight line) for two different fillings of the lowest band ($\nu = 1/4,\,1/2$). The blue dots represent the extrapolation to the thermodynamic limit of DMRG results for finite size Creutz ladders with periodic boundary conditions.}
\end{figure}


A change in the ground state energy $E_0(\Phi_i)$ as a function of the magnetic flux signals the presence of a persistent current flowing along the corresponding noncontractible loop, or equivalently the presence of propagating states that extend throughout the whole system. Indeed the response of the system to a change in the boundary conditions is the most fundamental way to discriminate between conductive and insulating states of matter~\cite{Resta:2011}.

\begin{figure}
\includegraphics{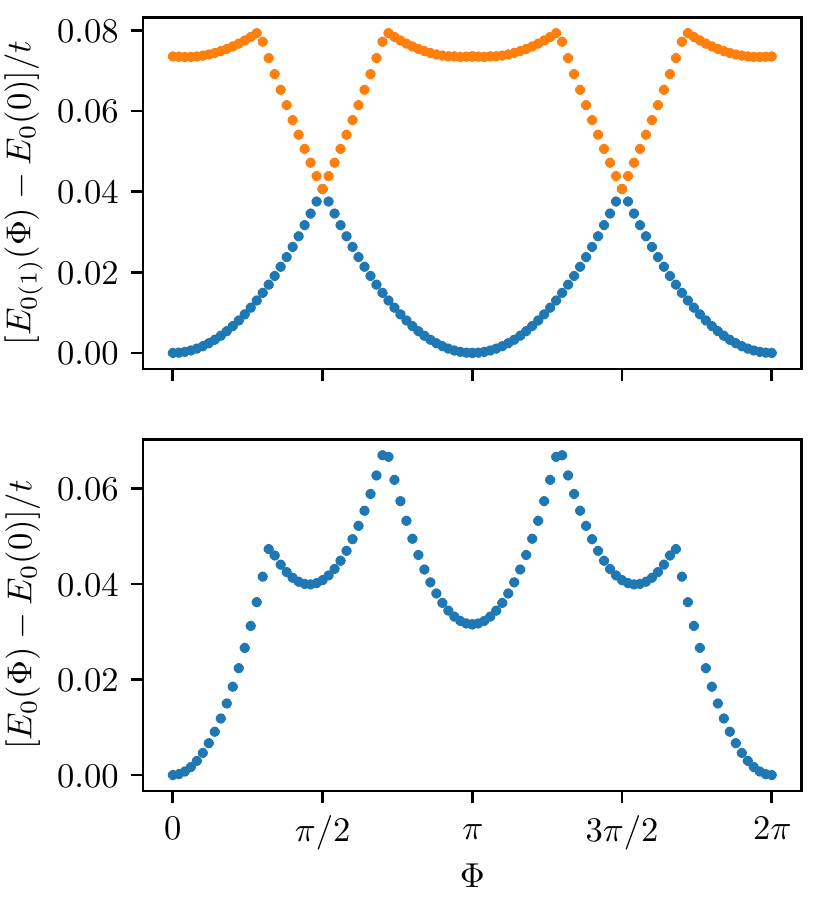}
\caption{\label{fig:Creutz_num_res_1}Energy difference $E_0(\Phi)-E_0(0)$ as a function of the magnetic flux for the Creutz ladder with spin-$1/2$ fermions and Hubbard interaction. These numerical data have been obtained using DMRG with periodic boundary conditions. \textit{Top panel}: the hopping matrix elements are as in Fig.~\ref{fig:lattice_models}\textbf{a}, corresponding to the perfectly flat band structure ($t_1 = t_2 = t$ in the notation of Appendix~\ref{app:def_Creutz_ladder}). The interaction strength is $U = -t < 0$ and $N_{\rm c} = 8$, $N_\uparrow = N_\downarrow = 4$. Both the ground state energy (blue dots) and the first excited state energy (orange dots) are plotted. Note the level crossing which ensures the $\pi$-periodic behavior of the energy.
\textit{Bottom panel:} the value of the hopping matrix elements has been changed to lift band flatness ($t_1=t$ and $t_2=0.6t$, in the notation of Appendix~\ref{app:def_Creutz_ladder}). As a consequence, $E_0(\Phi)-E_0(0)$ is only $2\pi$-periodic, but not $\pi$-periodic. For this plot $U = -t$, $N_{\rm c} = 10$, $N_\uparrow = N_\downarrow = 5$.}
\end{figure}

For small values of the magnetic flux the current intensity is measured by the phase stiffness~\cite{Fisher:1973}, also called superfluid or Drude weight $D_{\rm s}$~\cite{Scalapino:1992,Scalapino:1993}. Technically the superfluid weight and Drude weight are distinct quantities, but they coincide at zero temperature. This distinction is not important for what follows. The superfluid weight tensor is obtained from the general result
\begin{equation}
[D_{\rm s}]_{i,j} = \left.\frac{1}{V\hbar^2}\frac{\partial^2 \Omega}{\partial q_i\partial q_j}\right|_{\vc{q}=0}\,,
\end{equation}
where $\Omega$ is the thermodynamic grand potential and $V$ the volume (or area or length) of the system. The wavevector $\vc{q}$ has been introduced in Eq.~(\ref{eq:flux_insertion}) and~(\ref{eq:gauge_flux_insertion}). At zero temperature the derivative on the right hand side is equal to the derivative of the ground state energy 
\begin{equation}\label{eq:Ds_E0}
\left.\frac{\partial^2\Omega}{\partial q_i\partial q_j}\right|_{\vc{q}=0,T = 0} = \left.\frac{\partial^2E_0(\Phi_i)}{\partial q_i\partial q_j}\right|_{\vc{q}=0}\,.
\end{equation}
The superfluid weight of an isolated flat band with an attractive ($U < 0$) Hubbard interaction 
\begin{equation}\label{eq:Hubbard_int}
\mathcal{\hat H}_{\rm int} = U\sum_{\vc{i}\alpha}\hat{c}^\dagger_{\vc{i}\alpha\uparrow}\hat{c}_{\vc{i}\alpha\uparrow}\hat{c}^\dagger_{\vc{i}\alpha\downarrow}\hat{c}_{\vc{i}\alpha\downarrow}
\end{equation}
has been calculated analytically in a previous work by some of the authors~\cite{Peotta:2015}
\begin{equation}\label{eq:exact_Ds}
[D_{\rm s}]_{i,j} = \frac{4n_\phi |U|\nu(1-\nu)}{(2\pi)^d\hbar^2}\int_{\rm B.Z.}d^d\vc{k}\,\mathrm{Re}\, \mathcal{B}_{ij}(\vc{k})\,.
\end{equation}
Here $\nu$ is the flat band filling, $n_{\phi}^{-1}$ the number of orbitals on which the flat band wavefunction is nonvanishing, while the Quantum Geometric Tensor $\mathcal{B}_{ij}(\vc{k})$ is defined by
\begin{equation}\label{eq:QGT}
\mathcal{B}_{ij}(\vc{k}) = 2\bra{\partial_{k_i}g(\vc{k})}\big(1-\ket{g(\vc{k})}\bra{g(\vc{k})}\big)
 \ket{\partial_{k_j}g(\vc{k})}\,,
\end{equation}
with $\ket{g(\vc{k})}$ the periodic Bloch functions of the flat band. The periodic Bloch functions are defined in Sec.~\ref{sec:interpretation}. The real part $\mathrm{Re}\, \mathcal{B}_{ij}(\vc{k})$ of the Quantum Geometric Tensor which enters in Eq.~(\ref{eq:exact_Ds}) is known as the quantum metric. For details on the result~(\ref{eq:exact_Ds})-(\ref{eq:QGT}) see Refs.~\cite{Peotta:2015,Tovmasyan:2016,Liang:2017a,Liang:2017b}.

We use DMRG to obtain the ground state energy as a function of the magnetic flux $E_0(\Phi)$ for the Creutz ladder with twisted boundary conditions. Our
DMRG simulations are performed with the ALPS libraries~\cite{Bauer:2011,Dolfi:2014}. By fitting a quadratic function for small values of $\Phi$ we obtain a numerical estimate of the Drude weight which compares very well with Eq.~(\ref{eq:exact_Ds}) for $|U| \lesssim 4t = E_{\rm g}$, as shown in Fig.~\ref{fig:Drude_Creutz} for two distinct values of the filling ($\nu = 1/4,\,1/2$). The data points shown in Fig.~\ref{fig:Drude_Creutz} are obtained by finite size scaling of the Drude weight of Creutz ladders with different number of unit cells. Eq.~(\ref{eq:exact_Ds}) can be understood as a first order perturbative result where the expansion parameter is the interaction strength over the band gap $|U|/E_{\rm g}$~\cite{Tovmasyan:2016}, therefore it is exact in the isolated flat band limit. The first order result is accurate in a remarkably large range of values of $U$. A similar finite size scaling analysis confirming the validity of Eq.~(\ref{eq:exact_Ds}) has been performed also in a recent independent work~\cite{Mondaini:2018}.

The results of Fig.~\ref{fig:Drude_Creutz} serve as a benchmark of our DMRG simulations known to be more difficult in the case of periodic boundary conditions with respect to open ones~\cite{Rossini:2011}.
In the following we focus on the behavior of $E_0(\Phi)$ for large values of $\Phi$ rather than on the Drude weight. Moreover, we consider the case of the attractive Hubbard interaction for spin-$1/2$ fermions~(\ref{eq:Hubbard_int}) in our numerical simulations, thus Eq.~(\ref{eq:exact_Ds}) applies to the lattice models in Fig.~\ref{fig:lattice_models} which are time-reversal invariant, namely those with purely real hopping matrix elements. All of the bands of such lattice models have a nonzero quantum metric which means that the ground state has nonzero Drude/superfluid weight and the system is conductive in the presence of interactions. This result is important for what follows.
 
\section{Numerical results}
\label{sec:numerics}

It is evident, for example from Eq.~(\ref{eq:flux_insertion_boundary_1})-(\ref{eq:flux_insertion_boundary_2}), that the energy as a function of the magnetic flux is a $2\pi$-periodic function $E_0(\Phi) = E_0(\Phi+2\pi)$. In our units, where the magnetic flux is $\Phi_0 = hc/e = 2\pi$, this signals that the current carriers are have charge $e$, that is they are single quasiparticles. In fact for a lattice Hamiltonian with completely flat spectrum the energy is independent of the magnetic flux in absence of interactions. This is intuitively clear as in a flat band the group velocity is vanishing and steady state transport is not possible at all in absence of interactions, even at finite temperature. This is not the case for time-dependent transport~\cite{Wang:2017}. Another general property of the energy is to be an even function of flux $E_0(\Phi) = E_0(-\Phi)$ in the presence of time-reversal symmetry.

Instead the energy $E_0(\Phi)$ being a $\pi$-periodic function of the flux $E_0(\Phi) = E_0(\Phi+\pi)$ is an indication that the persistent ground state current is the result of the motion of composite particles with charge $2e$, that is Cooper pairs. The first main result of this work is that quite generally the ground state energy is a $\pi$-periodic function of the flux in the presence of interactions if the hopping parameters are tuned to realize a perfectly flat band structure. We systematically observe this behavior in the lattice models of Fig.~\ref{fig:lattice_models}. 

In the following we present numerical results for spin-$1/2$ fermions with Hamiltonians of the form 
\begin{equation}\label{eq:full_Ham}
\mathcal{\hat H}(\Phi_i) = \mathcal{\hat H}_{0,\uparrow}(\Phi_i)+\mathcal{\hat H}_{0,\downarrow}(\Phi_i)+\mathcal{\hat H}_{\rm int}\,.
\end{equation}
The noninteracting part of the Hamiltonian $\mathcal{\hat H}$ is composed of two identical copies $\mathcal{\hat H}_{0,\sigma = \uparrow,\downarrow}$ of one of the lattice models shown in Fig.~\ref{fig:lattice_models}, one copy for each component of the spin. The field operators are then labelled also by a spin index $\sigma$ ($\hat{c}_{\vc{i}\alpha} \to \hat{c}_{\vc{i}\alpha\sigma}$ in Eq.~(\ref{eq:free_H})). The magnetic fluxes $\Phi_{i = 1,2}$ enter in the boundary conditions of the noninteracting term as explained in Sec.~\ref{sec:boundary_condition}. The interaction term is the Hubbard interaction of Eq.~(\ref{eq:Hubbard_int}).

All lattice models in Fig.~\ref{fig:lattice_models} are time-reversal invariant with the exception of the 1D dice lattice for $k_2 \neq 0,\pi$. For such models Eq.~(\ref{eq:exact_Ds}) and~(\ref{eq:QGT}) imply that $E_0(\Phi)$ is not constant as a function of magnetic flux for $U < 0$, indicating a conductive ground state. A non-constant ground state energy as a function of magnetic flux is indeed what we observe in our numerical results. Our previous results on the superfluid weight of lattice models with flat bands guarantee that the system is superfluid in the thermodynamic limit~\cite{Peotta:2015,Julku:2016,Tovmasyan:2016,Liang:2017a,Liang:2017b}.

Whereas we consider only spin-$1/2$ fermions with the attractive Hubbard interaction in our numerical simulations, a case particularly relevant for solid state systems and current ultracold gas experiments, the analytical results presented here are generic for a large class of interaction terms and also for bosonic particles, as explained in Sec.~\ref{sec:analytics}.

\begin{figure}
\centering

\includegraphics[]{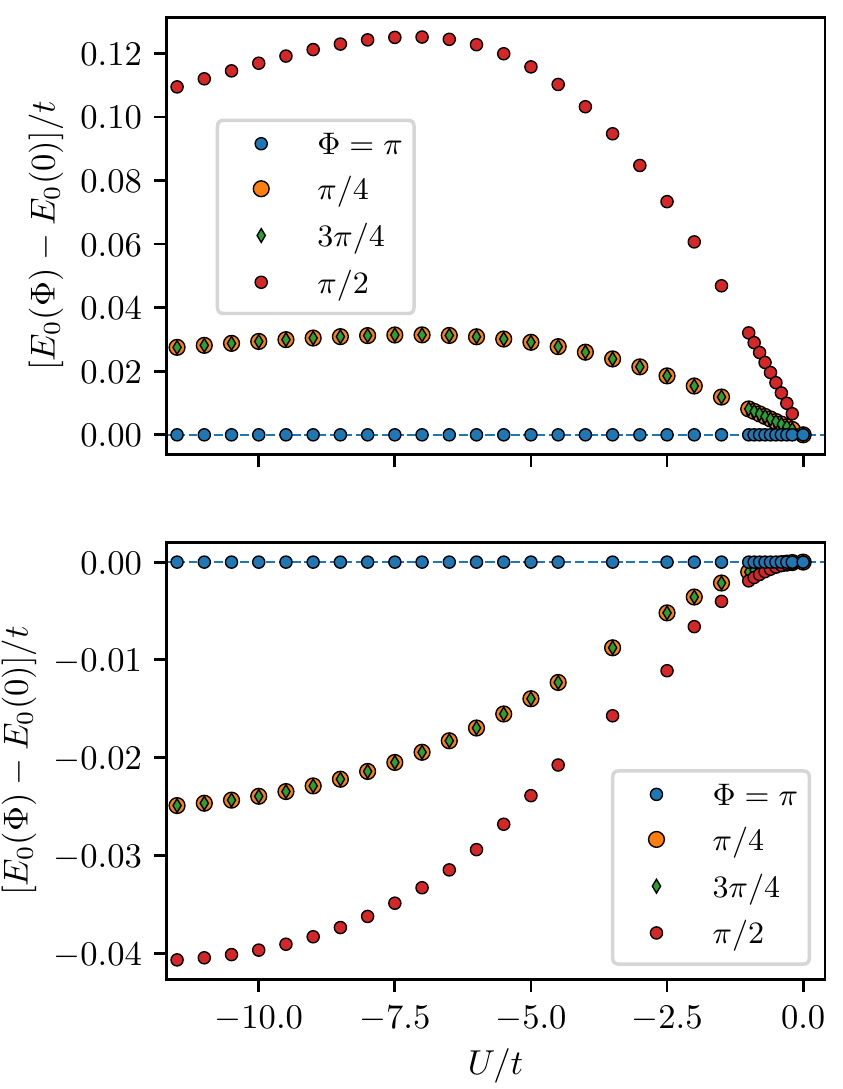} 
\caption{\label{fig:Creutz_num_res_2}Energy difference $[E_0(\Phi)-E_0(0)]/t$ for several values of the magnetic flux $\Phi$ as a function of the interaction strength $U$ in the Creutz ladder with perfectly flat bands, the same model considered in the top panel of Fig.~\ref{fig:Creutz_num_res_1}. In the top panel the length of the Creutz ladder is $N_{\rm c} = 10$ and the number of particles is $N_\uparrow = N_\downarrow = 5$ (spin balanced case); bottom panel $N_{\rm c} = 10$, $N_\uparrow = 6$, $N_\downarrow = 5$ (spin imbalanced case). 
}
\end{figure}

\subsection{Creutz ladder and diamond chain}
\label{sec:numerics_Creutz_diamond}
 
We first present numerical results for the Creutz ladder obtained using DMRG, as in Fig.~\ref{fig:Drude_Creutz}. 
In the top panel of Fig.~\ref{fig:Creutz_num_res_1} the ground state energy $E_0(\Phi)-E_0(0)$ is plotted as a function of $\Phi \in [0,2\pi]$ for $U = -t <0$. In the presence of a finite interaction the function $E_0(\Phi)$ is not constant, signalling that a persistent current is present in the ring for general twisted boundary conditions. The most important observation is that the energy difference is zero within numerical accuracy if $\Phi = \pi$, when the noninteracting Hamiltonian $\mathcal{\hat H}_0$ has a completely flat spectrum, as anticipated. On the other hand, if the bands are not flat as in the bottom panel of Fig.~\ref{fig:Creutz_num_res_1} the $\pi$-periodicity is lost.
This is a strong evidence that single-particle transport cannot occur at all if the bands are completely flat even in the presence of interactions. The energy of the first excited state $E_1(\Phi)-E_0(0)$ is also plotted in the top panel of Fig.~\ref{fig:Creutz_num_res_1}, showing that the $\pi$-periodic behavior is a consequence of the level crossing between the ground state and the first excited state.

The $\pi$-periodicity of the ground state energy shown in Fig.~\ref{fig:Creutz_num_res_1} is a robust phenomenon which does not depend on the interaction strength and also on the spin imbalance.
Indeed, in Fig.~\ref{fig:Creutz_num_res_2} the energy difference $E_0(\Phi)-E_0(0)$ is shown for several values of $\Phi = \pi/4,\,\pi/2,\,3\pi/4,\,\pi$ as a function of $U$. 
We see that $E_0(\Phi= \pi)-E_0(0)= 0$ independently of the value of $U$ and the number of spin-up and spin-down particles $N_\uparrow,\,N_{\downarrow}$.
This is particularly striking for the spin imbalanced case shown in the bottom panel of Fig.~\ref{fig:Creutz_num_res_2}.
In the case of spin imbalance one would expect that unpaired particles are present in the system and their motion would be reflected in the lifting of the $\pi$-periodicity of the ground state energy. According to the results of Fig.~\ref{fig:Creutz_num_res_2}, this does not happen and one may conclude that the unpaired particles remain localized even in the presence of interactions. This picture is correct as shown rigorously in Sec.~\ref{sec:analytics}.

\begin{figure}
\includegraphics[]{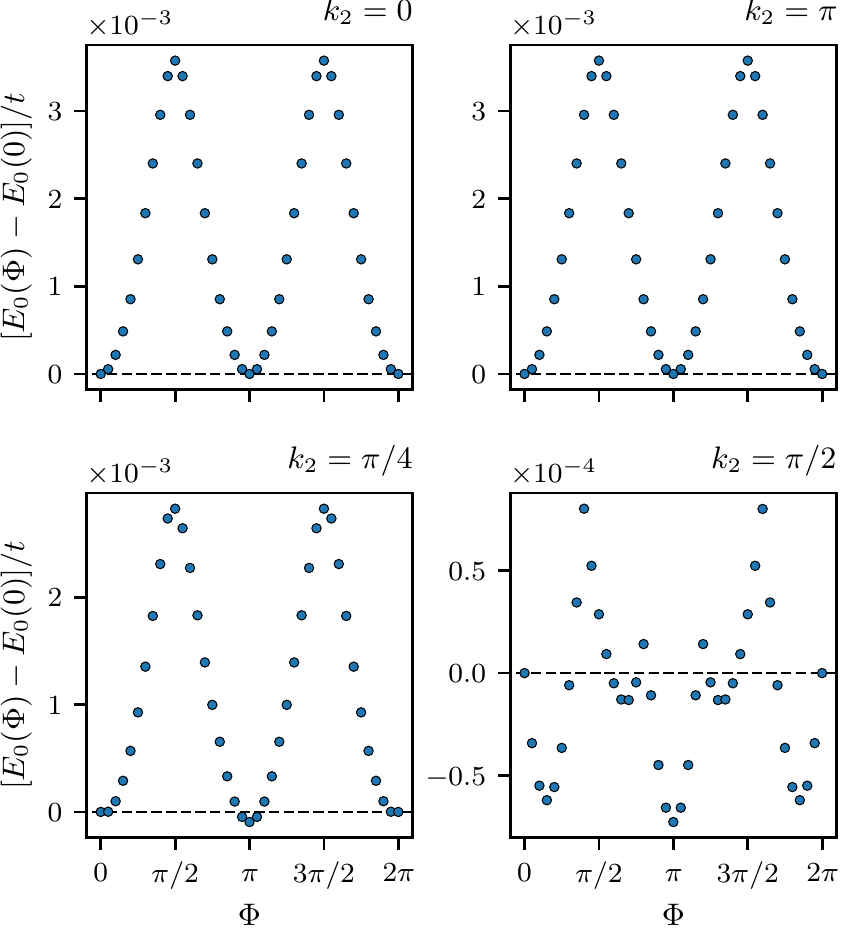}
\caption{\label{fig:1D_dice} Ground state energy as a function of magnetic flux for the 1D dice lattice of Fig.~\ref{fig:lattice_models}\textbf{c}. These numerical results have been obtained on a cluster of $N_{\rm c} = 6$ unit cells. In all four plots $U = -t <0$, $\varepsilon_{\rm h}=0$ (see Appendix~\ref{app:2D_dice_Ham}), $N_{\uparrow} = 3$, $N_\downarrow = 2$, while the value of the parameter $k_2$ (see Fig.~\ref{fig:lattice_models}\textbf{c}) is varied as indicated above each plot. Notice the breaking of $\pi$-periodicity for $k_2 = \pi/4,\,\pi/2$ in the bottom panels, which does not occur for $k_2 = 0,\,\pi$ as shown in the upper panels.}
\end{figure}

However, there are important differences between the cases with or without spin imbalance. One difference is that when $\Phi \neq \Phi_0/2$ the energy difference grows linearly in the spin balanced case (top panel in Fig.~\ref{fig:Creutz_num_res_2}) for small $|U|$ ($|E_0(\Phi)-E_0(0)|\propto |U|$), while the grow is quadratic in the spin imbalanced case (bottom panel). In fact from the analysis in our previous work~\cite{Tovmasyan:2016} we expect $|E_0(\Phi)-E_0(0)|\propto U^2/E_{\rm g}$, meaning that transport in the imbalance case can occur only through interband coupling induced by the interaction term, since the unpaired particles completely suppress transport at first order in $U$, that is at the level of the Hamiltonian projected on the lowest band. This point is discussed extensively in Sec.~\ref{sec:projected_conserved_quantities}. This is also reflected in the fact that $|E_0(\Phi)-E_0(0)|$ is smaller in the case of spin imbalance.

Fig.~\ref{fig:Creutz_num_res_2} shows that the energy is an even function also with $\Phi = \pi/2$ as center of inversion $E_0(\pi/2+\Phi) = E_0(\pi/2-\Phi)$. We have observed also in other numerical results not shown here that $E_0(\pi/2)\geq E_0(0)$ if the total particle number is even, while the opposite occurs if the total particle number is odd, as shown in the bottom panel of Fig.~\ref{fig:Creutz_num_res_2}. This means that the true ground state of the system would correspond to nonzero magnetic flux $\Phi = \pi/2$ for odd particle number, if the magnetic field were a dynamical degree of freedom and not a fixed external field as in our case. 

For the diamond chain the numerical results are qualitatively the same as for the Creutz ladder, that is for a perfectly flat band structure the ground state energy is $\pi$-periodic within numerical accuracy regardless of the interaction strength and the spin imbalance. Therefore we do not present here numerical results for the diamond chain. For both the Creutz ladder and the diamond chain we provide in Sec.~\ref{sec:analytics} rigorous analytical arguments which explain these observations.

\begin{figure}
\includegraphics{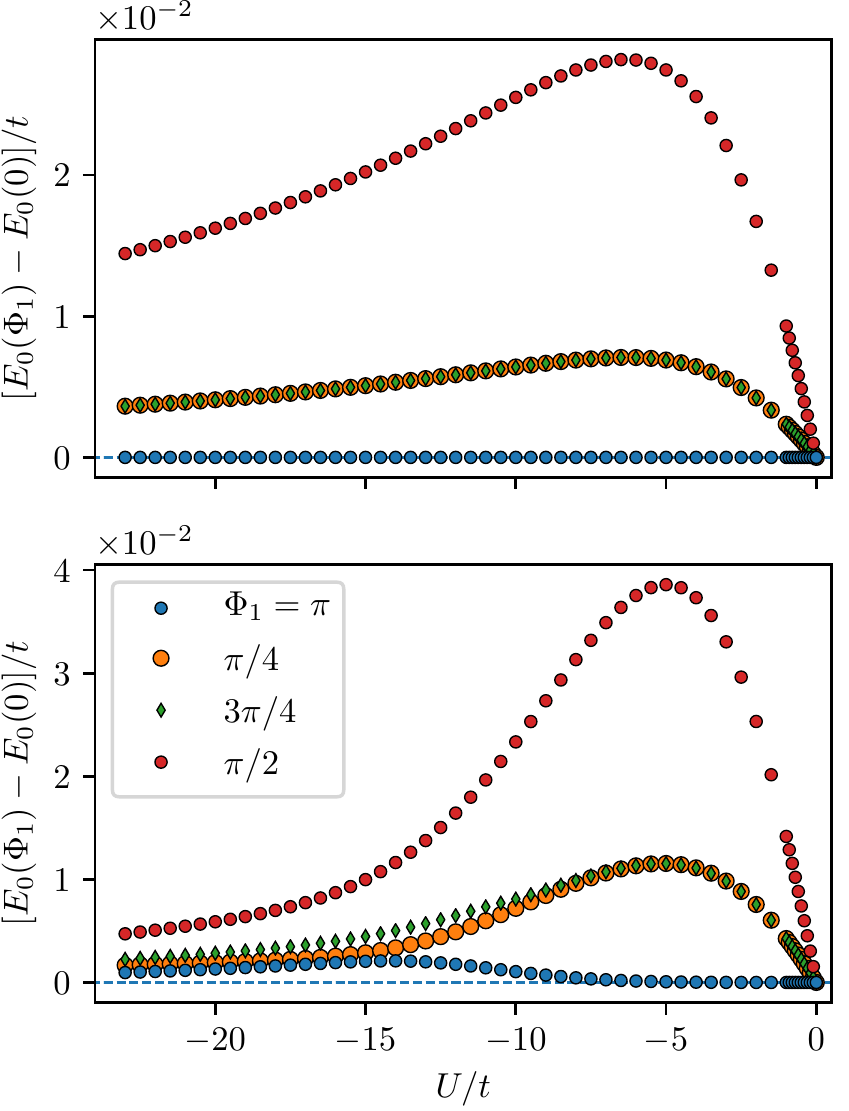}
\caption{\label{fig:2D_dice} Energy difference $E_0(\Phi_1)-E_0(0)$ in the 2D dice lattice of Fig.~\ref{fig:lattice_models}\textbf{b} with $\varepsilon_{\rm h}=0$ (see Appendix~\ref{app:2D_dice_Ham}) as a function of $U$ for various values of the magnetic flux $\Phi_1$. In the top panel ED results are shown for a cluster of size $6 \times 2$ ($N_1 \times N_2$) in the spin balanced case $(N_\uparrow = N_{\downarrow} = 2)$. In the bottom panel results are shown for the spin imbalanced case $(N_\uparrow = 3\,,N_\downarrow = 2)$ and cluster size $4 \times 2$. Note the small breaking of $\pi$-periodicity $E_0(\Phi_1 = \pi)\neq E_0(\Phi_1 = 0)$ in the spin imbalanced case. }
\end{figure}

\subsection{Dice lattice}
\label{sec:dice_numerical}

The previous analysis can be performed for the dice lattice as well.
We consider first the one dimensional reduction of the dice lattice presented in Fig.~\ref{fig:lattice_models}\textbf{c}.
In Fig.~\ref{fig:1D_dice} we provide numerical results for the 1D dice lattice obtained with ED on a cluster composed of $N_{\rm c} = 6$ unit cells. For ED we use a code which takes advantage of Graphical Processing Units~\cite{Siro:2012}. We consider from the outset the spin imbalanced case, that is $N_\uparrow = 3$ and $N_{\downarrow} = 2$. In the 1D dice lattice there is a free parameter, the phase factor $e^{ik_2}$ in Fig.~\ref{fig:lattice_models}\textbf{c}, that can be changed without lifting the perfect flatness of the band structure. In fact it is possible to introduce more parameters with the same property, as explained in Appendix~\ref{app:2D_dice_Ham}, but for our purposes the model shown in Fig.~\ref{fig:lattice_models}\textbf{c} is general enough.

In the spin balanced case the energy is always a $\pi$-periodic function of the magnetic flux for all values $k_2$ considered in our numerical simulations, as in the two lattice models analyzed previously in Sec.~\ref{sec:numerics_Creutz_diamond}. For this reason no numerical results in the case $N_\uparrow = N_\downarrow$ for the 1D dice lattice are shown here. On the other hand in the spin imbalanced case one observes a breaking of the $\pi$-periodicity of the ground state energy when $k_2 \neq 0,\pi$, see the two bottom plots in Fig.~\ref{fig:1D_dice}. This does not occur for the two special values $k_2 = 0,\,\pi$ for which the ground state energy is $\pi$-periodic within numerical accuracy. Precisely for these two special values we are able to provide an exact analytical argument in the same way as for the Creutz ladder and the diamond chain. This argument is explained in Sec.~\ref{sec:analytics} and shows for the 1D dice lattice with $k_2 = 0,\pi$ that the ground state energy is exactly $\pi$-periodic regardless of spin imbalance. On the other hand we have not been able to provide a similar argument for generic values of $k_2$. In fact Fig.~\ref{fig:1D_dice} shows that, if such argument exists, it must take into account spin imbalance.
Whether the methods used in Sec.~\ref{sec:analytics} can be extended in this direction is an open question.

\begin{figure}
\includegraphics{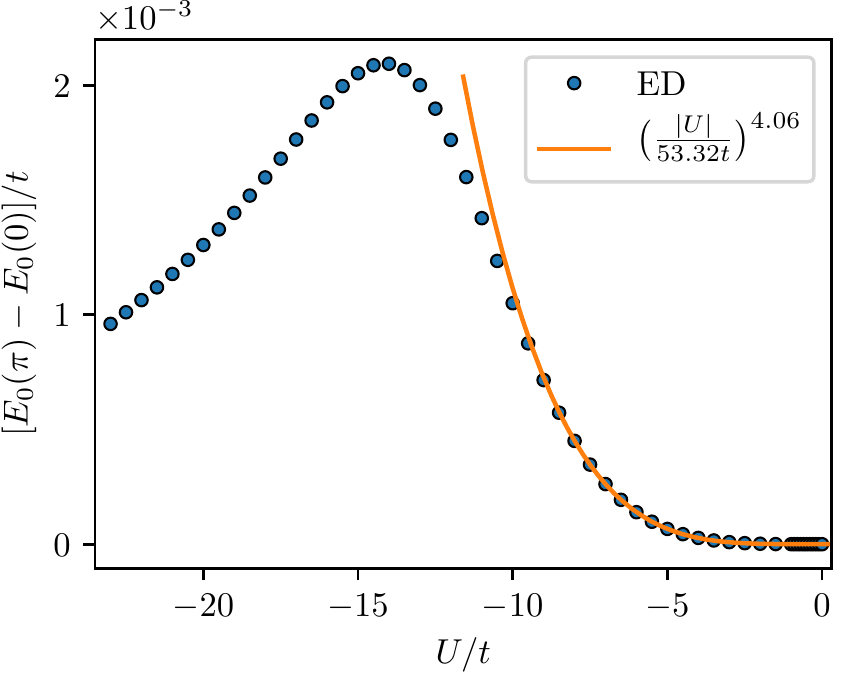}
\caption{\label{fig:fitting} Fit of the energy difference $E_0(\Phi_1 = \pi)-E_0(\Phi_1 = 0)$ in the 2D dice lattice. The blue dots correspond to the same numerical results shown in the bottom panel of Fig.~\ref{fig:2D_dice} for $\Phi_1 = \pi$. The energy difference is proportional to $(|U|/E_{\rm g})^4$ for small $U$ indicating that the loss of $\pi$-periodicity is an effect of interband coupling due to interactions.}
\end{figure}

Instead of providing numerical results analogous to Fig.~\ref{fig:Creutz_num_res_2} also for the 1D dice lattice, we consider the 2D dice lattice, where the behavior is substantially the same. The 1D dice lattice Hamiltonian is time reversal invariant precisely when $k_2 = 0,\,\pi$ since all the hopping matrix elements are real, while it is not time-reversal invariant for generic values of $k_2$ if the noninteracting part of the Hamiltonian is the same for both spin components. One may restore time-reversal invariance by taking $\mathcal{\hat H}_{0,\downarrow}$ to be the time-reversal conjugate of $\mathcal{\hat H}_{0,\uparrow}$ for $\Phi_i = 0$ in Eq.~(\ref{eq:full_Ham}). However we have chosen not to do so in this work. Indeed, the ED results of Fig.~\ref{fig:2D_dice} for the 2D dice lattice of Fig.~\ref{fig:lattice_models}\textbf{b} show that the breaking of $\pi$-periodicity can occur even when the Hamiltonian is time-reversal invariant in the above sense. Also for the 2D dice lattice the ground state energy is always $\pi$-periodic in the spin balanced case, while it is not so in the presence of spin imbalance, as for the 1D dice lattice. From Fig.~\ref{fig:2D_dice} it is apparent that the nonzero splitting $E_0(\Phi_1 = \pi)-E_0(\Phi_1 = 0)$ in the spin imbalanced case is not linear in the coupling $U$ (compare with the data for $\Phi_1 \neq \pi$). Indeed, it is shown by the fit in Fig.~\ref{fig:fitting} that $E_0(\Phi_1 = \pi)-E_0(\Phi_1 = 0) \propto E_{\rm g}(U/E_{\rm g})^4$ for small $U$, with good accuracy. This clearly indicates that the breaking of $\pi$-periodicity is not an intrinsic effect associated to the flat band but rather an effect of the interband coupling due to interactions. In particular the many-body Hamiltonian projected on the flat band subspace does show $\pi$-periodicity and the breaking thereof can be recovered only at a rather high order in a perturbative expansion in the small parameter $|U|/E_{\rm g}$. See Ref.~\cite{Tovmasyan:2016} and Sec.~\ref{sec:projected_conserved_quantities} for more details on how the perturbative expansion can be carried out in practice.
Figs.~\ref{fig:1D_dice}-\ref{fig:2D_dice}-\ref{fig:fitting} suggest that even in a lattice Hamiltonian with a completely flat band structure, propagating states for unpaired quasiparticles can arise as a consequence of interband coupling. The exact nature of these quasiparticle states is an intriguing question which might be the subject of future investigations. However it would be necessary to first exclude  that the splitting $|E_0(\pi)-E_0(0)| \neq 0$ observed in Fig.~\ref{fig:1D_dice}-\ref{fig:2D_dice}-\ref{fig:fitting} is a finite size effect which vanishes in the thermodynamic limit.

For small values of the coupling $U$ the interband mixing is negligible and all the dynamics occurs in the flat band subspace. In this limit we find again evidence that transport can occur only through the correlated motion of pairs of particles, while single quasiparticles are localized.

\section{Analytical results}
\label{sec:analytics}

\subsection{Graph automorphisms and intertwining operators}
\label{sec:graph_automorphism}

In this section we provide a simple analytic argument rigorously proving that the ground state energy as a function of magnetic flux $E_0(\Phi)$ is $\pi$-periodic in the case of the one dimensional lattices considered here. As anticipated we explicitly construct the intertwining unitary operator $\mathcal{\hat U}$ which performs the flux insertion~(\ref{eq:intertwining_op}). It turns out that in all cases $\mathcal{\hat U}$ is a canonical tranformation of the field operators and it is associated to symmetries of the lattice considered as a graph.

It is useful to introduce few standard definitions in graph theory. A graph $G$ is defined by a set of vertices $V$ and a set of edges $E$ whose elements are unordered pairs $(v,w)$ of distinct vertices. Two vertices $v,w\in V$ are said to be connected, or adjacent, if $(v,w)\in E$. The lattices shown in Fig.~\ref{fig:lattice_models} are all examples of graphs, if the information encoded in the edge color is neglected. The vertices are the lattice sites. A \textit{graph automorphism} is a permutation of the vertices $\sigma: \, V\to V$ which preserves the adjacency relations, that is $(\sigma(v),\sigma(w)) \in E$ if and only if $(v,w)\in E$. The set of all graph automorphisms is a group under composition, called the \textit{graph automorphism group}. The graph automorphism group contains as a subgroup the space group of the lattice, that is the group of rigid transformations, such as translations and reflections. However it generally contains also transformations that do not belong to the space group. These are precisely the ones we are interested in.

\begin{figure}
\includegraphics[scale = 0.9]{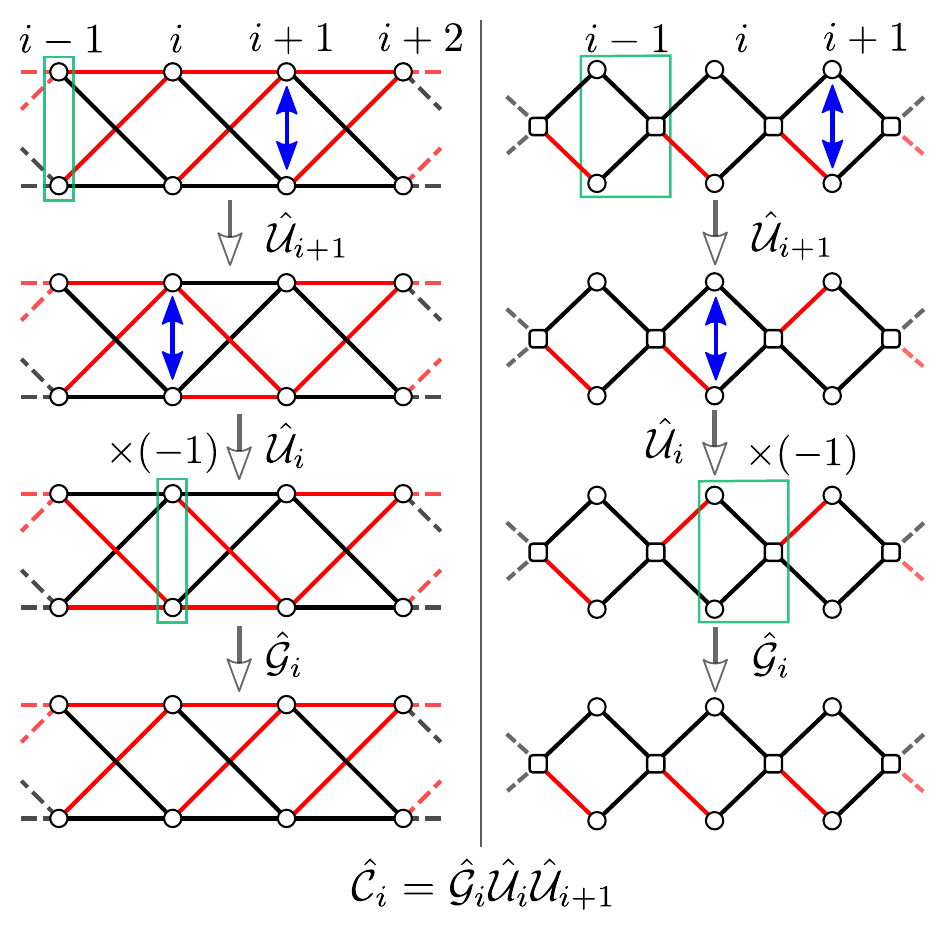}
\caption{\label{fig:intertwining} Intertwining operators and associated conserved quantities for the Creutz ladder and the diamond chain. The sign of the hopping matrix elements is encoded in the edge colors as in Fig.~\ref{fig:lattice_models}. The permutations of the orbitals (graph automorphisms) associated to the intertwining operators $\mathcal{\hat U}_i$ are indicated by blue arrows. In both the Creutz ladder and the diamond chain, $\mathcal{\hat U}_i$ performs a permutation of a pair of field operators inside unit cell $i$, see Eqs.~(\ref{eq:U_creutz_1})-(\ref{eq:U_creutz_2}) for the Creutz ladder and Eqs.~(\ref{eq:U_diam_2})-(\ref{eq:U_diam_3}) for the diamond chain. After applying the operator $\mathcal{\hat U}_{i+1}$ all the hopping terms between unit cell $i$ and $i+1$ take a minus sign, as shown graphically in the first step in the figure. This operation amounts to the insertion of a $\Phi = \pi$ flux through the ring according to Fig.~\ref{fig:flux_insertion}\textbf{d}, namely $\mathcal{\hat U}_{i+1}\mathcal{\hat H}(\Phi = 0)\mathcal{\hat U}_{i+1}^\dagger = \mathcal{\hat H}(\Phi = \pi)$. The subsequent action of $\mathcal{\hat U}_{i}$ adds again $\pi$ to the total flux through the ring (second step). A total flux $\Phi = 2\pi$ is equivalent to zero flux since $\mathcal{\hat H}(\Phi = 0)$ and $\mathcal{\hat H}(\Phi = 2\pi)$ are related by the gauge transformation $\mathcal{\hat G}_i$ (\ref{eq:G_Creutz_def}), as shown in the last step. The gauge transformation $\mathcal{\hat G}_i$ multiplies all the field operators inside unit cell $i$ by $-1$ (see third row in the figure). Therefore $\mathcal{\hat C}_i = \mathcal{\hat G}_i\mathcal{\hat U}_{i}\mathcal{\hat U}_{i+1}$ is a local conserved quantity of the Hamiltonian in the presence of a large class of interaction terms. The same graphical approach 
works also for the 1D dice lattice (Fig.~\ref{fig:lattice_models}\textbf{c}) with $k_2 = 0,\pi$ (not shown here).}
\end{figure}

The Creutz ladder and the diamond chain provide the simplest examples of graph automorphisms that do not belong to the space group of the lattice. For the Creutz ladder this is the permutation $(i,\alpha = 1) \leftrightarrow (i,\alpha = 2)$ of the two vertices inside the $i$-th unit cell. The graph automorphism for the Creutz ladder is illustrated in Fig.~\ref{fig:intertwining}. This is a local transformation since it involves only two lattice sites and there are $N_{\rm c}$ independent transformations of this kind in a lattice with $N_{\rm c}$ unit cells. The intertwining operators for the Creutz ladder can be constructed simply by performing exactly the same permutation on the field operators
\begin{gather}
\mathcal{\hat U}_i\hat{c}_{i,1}\mathcal{\hat U}_i^\dagger = \hat{c}_{i,2}\,, \label{eq:U_creutz_1}\quad
\mathcal{\hat U}_i\hat{c}_{i,2}\mathcal{\hat U}_i^\dagger = \hat{c}_{i,1}\,,\\
\mathcal{\hat U}_i\hat{c}_{j\alpha}\mathcal{\hat U}_i^\dagger = \hat{c}_{j\alpha}\quad \text{for} \quad i \neq j. \label{eq:U_creutz_2}
\end{gather}
It is not necessary to provide a more explicit form for the unitary operators $\mathcal{\hat U}_i$ since Eqs.~(\ref{eq:U_creutz_1})-(\ref{eq:U_creutz_2}) completely specify their action on the whole Hilbert space. Any operator $\mathcal{\hat U}_i$ performs the flux insertion as required, as shown in Fig.~\ref{fig:intertwining} in a purely graphical manner by keeping track of the edge colors under the permutation of the sites of the lattice.


Obviously also the Hubbard interaction term in Eq.~(\ref{eq:Hubbard_int}) is invariant under the action of the intertwining operator (\ref{eq:U_creutz_1})-(\ref{eq:U_creutz_2}). In fact, in Eqs.~(\ref{eq:U_creutz_1})-(\ref{eq:U_creutz_2}) the spin index has been suppressed to emphasize that our argument works for particles with arbitrary spin. In the following it is understood that all intertwining operators in the case of spin-$1/2$ particles are in fact the product of identical operators for each spin component $\mathcal{\hat U}_i = \mathcal{\hat U}_{i,\uparrow}\mathcal{\hat U}_{i,\downarrow}$, where $\mathcal{\hat U}_{i,\sigma = \uparrow,\downarrow}$ are both defined by Eqs.~(\ref{eq:U_creutz_1})-(\ref{eq:U_creutz_2}) for the Creutz ladder or by similar definitions for the other models, to be presented below.
Therefore Eq.~(\ref{eq:intertwining_op}) is valid for the full many-body Hamiltonian~(\ref{eq:full_Ham}) with an Hubbard interaction. In fact our argument provides a simple characterization of all the interaction terms $\mathcal{\hat H}_{\rm int}$ for which Eq.~(\ref{eq:intertwining_op}) holds, which can be much more general that the Hubbard one~(\ref{eq:Hubbard_int}). In particular our results apply to the interacting Creutz model with spinless fermions studied in Ref.~\cite{Junemann:2017}. In agreement to our general results, single particle transport is enabled in this model only by a term explicitly breaking the orbital symmetry $\hat{V}_{\rm imb} = \frac{\Delta\varepsilon}{2}\sum_{i}
(\hat{c}_{i,1}^\dagger\hat{c}_{i,1}-
\hat{c}_{i,2}^\dagger\hat{c}_{i,2})$ (see Eq.~(5) in Ref.~\cite{Junemann:2017}), which lifts the band flatness, and is otherwise completely suppressed even for finite interactions (see for example Fig.~5 in Ref.~\cite{Junemann:2017}). In fact in the case of spinless fermions one can show that even pair transport and any kind of transport are completely suppressed if the interaction term is invariant under $\mathcal{\hat U}_i$. This argument will be provided in Sec.~\ref{sec:interpretation}. 

It is also straightforward to provide the intertwining operator for the diamond chain (Fig.~\ref{fig:lattice_models}\textbf{d}). The corresponding local graph automorphism is given by the permutation $(i,\alpha = 1)\leftrightarrow (i,\alpha = 2)$ for a given unit cell $i$ and it is shown in Fig.~\ref{fig:intertwining} as well. Then the operator $\mathcal{\hat U}_i$ is defined by
\begin{gather}
\mathcal{\hat U}_i\hat{c}_{i,1}\mathcal{\hat U}_i^\dagger = \hat{c}_{i,2}\,,\quad
\mathcal{\hat U}_i\hat{c}_{i,2}\mathcal{\hat U}_i^\dagger = \hat{c}_{i,1}\,,\label{eq:U_diam_1}\\
\mathcal{\hat U}_i\hat{c}_{i,3}\mathcal{\hat U}_i^\dagger = \hat{c}_{i,3}\,, \label{eq:U_diam_2}\\
\mathcal{\hat U}_i\hat{c}_{j\alpha}\mathcal{\hat U}_i^\dagger = \hat{c}_{j\alpha}\quad \text{for} \quad i \neq j. \label{eq:U_diam_3}
\end{gather}
Again one can graphically check that this intertwining operator performs the $\pi$-flux insertion (see Fig.~\ref{fig:intertwining} right).

Next we provide the intertwining operator for the 1D dice lattice of Fig.~\ref{fig:lattice_models}\textbf{c}. We consider only the two values of the parameter $k_2 = 0,\,\pi$ for which the Hamiltonian is time-reversal invariant. This lattice possesses two distinct families of local graph automorphisms, one is $(i,1)\leftrightarrow (i,2)$ and the other is $(i,5)\leftrightarrow (i,6)$. The intertwining operator for $k_2=0$ performs both permutations inside a single unit cell $i$
\begin{gather}
\mathcal{\hat U}_i\hat{c}_{i,1}\mathcal{\hat U}_i^\dagger = \hat{c}_{i,2}\,,\quad
\mathcal{\hat U}_i\hat{c}_{i,2}\mathcal{\hat U}_i^\dagger = \hat{c}_{i,1}\,, \label{eq:U_dice1_2}\\
\mathcal{\hat U}_i\hat{c}_{i,5}\mathcal{\hat U}_i^\dagger = \hat{c}_{i,6}\,, \quad
\mathcal{\hat U}_i\hat{c}_{i,6}\mathcal{\hat U}_i^\dagger = \hat{c}_{i,5}\,, \label{eq:U_dice1_3}\\
\mathcal{\hat U}_i\hat{c}_{i,3}\mathcal{\hat U}_i^\dagger = \hat{c}_{i,3}\,. \label{eq:U_dice1_4}\\
\mathcal{\hat U}_i\hat{c}_{i,4}\mathcal{\hat U}_i^\dagger = -\hat{c}_{i,4}\,. \label{eq:U_dice1_5}\\
\mathcal{\hat U}_i\hat{c}_{j\alpha}\mathcal{\hat U}_i^\dagger = \hat{c}_{j\alpha}\quad \text{for} \quad i \neq j. \label{eq:U_dice1_6}
\end{gather}
Eq.~(\ref{eq:U_dice1_5}) alone is just a gauge transformation and strictly speaking it is not needed. However it is convenient since after the flux-insertion transformation only the hopping matrix elements crossing the boundary between unit cells $i-1$ and $i$ are modified, as in Eq.~(\ref{eq:flux_insertion_boundary_1}) and Fig.~\ref{fig:flux_insertion}\textbf{d}.

The flux-insertion transformation for $k_2 = \pi$ is very similar, but distinct, since it involves permutations of sites on two adjacent unit cells
\begin{gather}
\mathcal{\hat U}_i\hat{c}_{i-1,5}\mathcal{\hat U}_i^\dagger = -\hat{c}_{i-1,6}\,,\quad
\mathcal{\hat U}_i\hat{c}_{i-1,6}\mathcal{\hat U}_i^\dagger = -\hat{c}_{i-1,5}\,, \label{eq:U_dice2_1}\\
\mathcal{\hat U}_i\hat{c}_{i-1,\alpha}\mathcal{\hat U}_i^\dagger = \hat{c}_{i-1,\alpha}\,,\quad \text{for}\quad \alpha = 1,2,3,4 \label{eq:U_dice2_2}\\
\mathcal{\hat U}_i\hat{c}_{i,1}\mathcal{\hat U}_i^\dagger = \hat{c}_{i,2}\,, \quad
\mathcal{\hat U}_i\hat{c}_{i,2}\mathcal{\hat U}_i^\dagger = \hat{c}_{i,1}\,, \label{eq:U_dice2_3}\\
\mathcal{\hat U}_i\hat{c}_{i\alpha}\mathcal{\hat U}_i^\dagger = \hat{c}_{i\alpha}\,,\quad \text{for}\quad \alpha = 3,4,5,6 \label{eq:U_dice2_4}\\
\mathcal{\hat U}_i\hat{c}_{j\alpha}\mathcal{\hat U}_i^\dagger = \hat{c}_{j\alpha}\quad \text{for} \quad i \neq j\quad \text{and} \quad i-1 \neq j\,. \label{eq:U_dice2_5}
\end{gather}
Again this transformation is designed in such a way that only the hopping matrix elements crossing the boundary between unit cells $i-1$ and $i$ change sign.

Whereas in all the lattices considered so far the intertwining operators can be readily constructed from graph automorphisms of the lattice, we have not been able to construct a local intertwining operator for generic values of $k_2 \neq 0,\pi$ in the 1D dice lattice. However it is possible to provide an intertwining operator for a 1D dice lattice with odd length $N_{\rm c} \mod 2 = 1$ acting simultaneously on orbitals $\alpha = 1,2,5,6$ in all unit cells (see Appendix~\ref{app:global_U}). 
A similar intertwining operator can be constructed for the Creutz ladder, again only if the length is odd. In the case of the Creutz ladder the intertwining operator exists even if the bands are not flat, which implies that the $\pi$-periodicity of the spectrum can occur even in absence of a perfectly flat band structure (see Appendix~\ref{app:global_U}). This may seem to cast doubt on our general approach of relating the periodicity of the Aharonov-Bohm effect to the charge of the carriers. However, as shown in Fig.~\ref{fig:Creutz_num_res_1}, the ground state energy is not $\pi$-periodic in a Creutz ladder of even length when the bands are not flat, in contrast to odd length. Moreover, the intertwining operators presented in Appendix~\ref{app:global_U} are not local since they act simultaneously on an extensive number of lattice sites and for this reasons they cannot be used to construct conserved quantities as explained in Sec.~\ref{sec:conserved quantities}. 
In order to avoid this even-odd effect, which we regard as accidental for our purposes, in Figs.~\ref{fig:Creutz_num_res_1}-\ref{fig:fitting} we have presented numerical results only for lattices of even length ($N_1\mod 2 = 0$) along the noncontractible loop through which the magnetic flux is varied, that is along direction $\vc{a}_1$ for the 2D dice lattice (see Appendix~\ref{app:2D_dice_Ham}).

The graph automorphism group of the full 2D dice lattice of Fig.~\ref{fig:lattice_models} coincides with its space group, which means that there are no local graph automorphisms available to construct the intertwining operators as in the 1D case. This intuitive fact can be checked by using one of the several numerical tools available to calculate the automorphism group of a generic graph. We have used the codes \textsc{nauty} and \textsc{traces} in this work~\cite{McKay201494}. As in the case of the 1D dice lattice for generic $k_2$, we provide for the 2D dice lattice a nonlocal intertwining operator which relates the Hamiltonians with $\Phi_1$ and $\Phi_1+\pi$ if the number of unit cells in the same direction $N_1$ is odd, see Appendix~\ref{app:global_U}.

As remarked in Sec.~\ref{sec:dice_numerical} an intertwining operator cannot be constructed in the 1D dice lattice for $k_2 \neq 0,\pi$ and in the 2D dice lattice since we observe a breaking of $\pi$-periodicity in the spin imbalanced case, as shown in Fig.~\ref{fig:1D_dice} and~\ref{fig:2D_dice}. An open question is whether this could be done only in the case $N_\uparrow = N_\downarrow$, for which we do observe $\pi$-periodicity in all the lattice Hamiltonians with completely flat band structure studied here.

\subsection{Conserved quantities}
\label{sec:conserved quantities}

The intertwining operators for the Creutz ladder~(\ref{eq:U_creutz_1})-(\ref{eq:U_creutz_2}), the diamond chain~(\ref{eq:U_diam_1})-(\ref{eq:U_diam_3}) and the 1D dice lattice for $k_2 = 0,\pi$~(\ref{eq:U_dice1_2})-(\ref{eq:U_dice2_5}) are local, that is they act nontrivially only on a small number of unit cells which is independent of the system size. In fact, one  has an extensive number of intertwining operators $\mathcal{\hat U}_i$ in each case, labelled by $i = 1,\dots,N_{\rm c}$. An external perturbation may break some of these but not all if the perturbation is local. As long as a single intertwining operator exists which satisfies Eq.~(\ref{eq:intertwining_op}) the $\pi$-periodicity property is preserved. A physical interpretation of this would be that, as long as only pairs of particles are allowed to cross even a small section of the lattice, then all the eigenvalues of the Hamiltonian are $\pi$-periodic as functions of the magnetic flux.

From our perspective the most important consequence of locality is the existence of an extensive number of mutually commuting integrals of motion or conserved quantities. These are the product of two distinct intertwining operators $\mathcal{\hat U}_i\mathcal{\hat U}_j$. This is a conserved quantity since, by using repeatedly Eq.~(\ref{eq:intertwining_op}),
\begin{equation}
\mathcal{\hat U}_i\mathcal{\hat U}_j \mathcal{\hat H}(\Phi)\mathcal{\hat U}^\dagger_j\mathcal{\hat U}_i^\dagger
=\mathcal{\hat U}_i\mathcal{\hat H}(\Phi+\Phi_0/2)\mathcal{\hat U}_i^\dagger = 
\mathcal{\hat H}(\Phi+\Phi_0)\,,
\end{equation} 
and $\mathcal{\hat H}(\Phi)$ and $\mathcal{\hat H}(\Phi+\Phi_0)$ are physically equivalent being related by a gauge transformation. It is convenient to include this gauge transformation $\mathcal{\hat G}$ in the definition of the conserved quantities. They are mutually commuting because also the intertwining operators commute with each other $\mathcal{\hat U}_i\mathcal{\hat U}_j = \mathcal{\hat U}_j\mathcal{\hat U}_i$. 
If $i\neq j$ then $\mathcal{\hat U}_i\mathcal{\hat U}_j$ is different from the identity, therefore it is a nontrivial local conserved quantity of the many-body Hamiltonian. All of these conserved quantities can be constructed by a smaller subset of independent ones of the form $\mathcal{\hat U}_i\mathcal{\hat U}_{i+1}$ with $i = 1,\dots,N_{\rm c}$ since the intertwining operator is idempotent in all cases ($\mathcal{\hat U}_i^2 = 1$). The number of independent conserved quantities scales with the system size, but is strictly less than the total number of degrees of freedom in all of the models considered here. In this sense these Hamiltonians are all at least partially integrable. The question of whether they are also fully integrable is an interesting one, but it will be not addressed here, although we will add a remark in Sec.~\ref{sec:projected_conserved_quantities}. 
The existence of these conserved quantities can be checked in a purely graphical way, as shown in Fig.~\ref{fig:intertwining} for the Creutz ladder and the diamond chain. A similar graphical method works also in the case of the 1D dice lattice for $k_2 = 0,\pi$ (not shown).

The only model for which these conserved quantities have been discussed previously in the literature is the diamond chain~\cite{Doucot:2002}. The fundamental building blocks of these conserved quantities, the local intertwining operators on the other hand have not been presented before in the literature. The approach based on the intertwining operators has the benefit to significantly enlarge the scope of the results of Ref.~\cite{Doucot:2002}. It is clear that the conserved quantities are present both for fermions and bosons since they are implemented by canonical transformations. In fact, being a simple permutation of the field operators modulo a gauge transformation, they preserve also the mixed commutation/anticommutation relations of hardcore bosons. Therefore they can be useful also for spin models that can be mapped to hardcore bosons, for example the quantum Ising model on the diamond chain studied in Ref.~\cite{Coester:2013}. In this last work local $\mathbb{Z}_2$ symmetries are found in the low energy effective Hamiltonian obtained perturbatively, but they were not related to symmetries of the full Hamiltonian, which in fact are present, as our general argument shows.

The class of interaction terms that allow for these conserved quantities is rather large and includes all of the density-density interactions terms that are symmetric under the graph automorphisms used to build the intertwining operator. The presence of an extensive number of local conserved quantities in the case of the Creutz ladder is a new result. This result is relevant for recent works on the Creutz ladder~\cite{Takayoshi:2013,Tovmasyan:2013,Tovmasyan:2016,Junemann:2017} where both bosonic and fermionic Hubbard models have been considered. We note also that in a recent work by Drescher and Mielke~\cite{Drescher:2017} similar local conserved quantities associated to graph automorphisms have been found in a lattice Hamiltonian where only some bands are perfectly flat. In this case the conserved quantities seem not to be constructed from intertwining operators that perform a $\Phi_0/2$-flux insertion. The application of graph theory concepts to Hubbard models with flat bands also goes back to Mielke~\cite{Mielke:1991,Mielke:1991b,Mielke:1992,Mielke:1993}. Interestingly, in a recent work~\cite{Roentgen:2018} it was shown that the existence of local permutations commuting with a generic quadratic lattice Hamiltonian implies that a flat band is present in the band structure. Again the underlying mathematical theory is borrowed from (spectral) graph theory.

An important question is the physical interpretation of these conserved quantities and why they seem to appear precisely when the noninteracting Hamiltonian has a flat spectrum. This is the subject of the next section where the conserved quantities are related to localized single quasiparticle excitations of the many-body Hamiltonian.

\subsection{Wannier functions and physical interpretation}
\label{sec:interpretation}

In this section we interpret the conserved quantities as the parity of the occupation number of localized single quasiparticle eigenstates of the many-body Hamiltonian. These eigenstates are associated to compactly localized Wannier functions which we now briefly discuss.

Due to the macroscopic degeneracy of a flat band there are different convenient bases of eigenfunctions that can be used. For our purposes the basis of Wannier functions is the most useful. The Wannier functions $w_n(\vc{j})$ are the Fourier transform of the Bloch functions $g_n(\vc{k})$ relative to a given band labelled by the band index $n=1,\dots,N_{\rm orb}$
\begin{equation}
\label{eq:Wannier_func_def}
w_n(\vc{j}) = \frac{V_{\rm c}}{(2\pi)^d}\int_{\rm B.Z.}\,d^d\vc{k}\, e^{i\vc{k}\cdot \vc{r}_{\vc{j}}} g_{n}(\vc{k})\,.
\end{equation}
Here $V_{\rm c}$ is volume of the unit cell in $d=3$, area in $d=2$ or length in $d=1$ (in the latter case $V_{\rm c} =a $ with $a$ the lattice spacing) and the integral is over the Brillouin zone (B.Z.). Both the Wannier functions and the Bloch function are vector-valued functions of $\vc{j}$ and $\vc{k}$, respectively, with $N_{\rm orb}$ components. The components are denoted by $w_n(\vc{j},\alpha)$ and $g_n(\vc{k},\alpha)$. The Bloch functions are eigenstates with energy $\varepsilon_n(\vc{k})$ of the Fourier transform of the hopping matrix
\begin{equation}
\widetilde{K}(\vc{k}) g_n(\vc{k}) = \varepsilon_n(\vc{k})g_n(\vc{k})\,.
\end{equation}
If the Bloch functions are normalized and orthogonal, that is if
\begin{equation}\label{eq:Bloch_normalization}
[g_n(\vc{k})]^\dagger g_{n'}(\vc{k}) = \delta_{n,n'}\,,
\end{equation}
then the Wannier functions and their translations form a complete orthonormal set
\begin{equation}\label{eq:Wannier_normalization}
\sum_{\vc{j}}[w_n(\vc{j}-\vc{l})]^\dagger w_{n'}(\vc{j}-\vc{l}')=\delta_{n,n'}\delta_{\vc{l},\vc{l}'}\,.
\end{equation}
Notice that in case of band degeneracies there is no unique choice of orthogonal Bloch functions. Even in absence of band degeneracies the condition~(\ref{eq:Bloch_normalization}) does not completely specify the Bloch function since it allows for multiplication by an arbitrary phase factor, that is $g_n(\vc{k})\to e^{i\phi(\vc{k})}g_n(\vc{k})$. This transformation leads in general to very different Wannier functions, in particular the smoother the Bloch functions are in momentum space (as functions of $\vc{k}$), the more localized the Wannier functions will be in real space, due to the general properties of the Fourier transform~\cite{Marzari:2012}. Therefore it is generally convenient to choose the phase factor in such a way that the Bloch functions are analytic throughout the Brillouin zone, since this leads to Wannier functions that decay exponentially at large distances. The Wannier functions are eigenstates of the single-particle Hamiltonian only if the band is flat ($\varepsilon_n(\vc{k})=\varepsilon_n$).

An interesting case is when the Wannier functions are localized (i.e. nonzero) on a finite number of unit cells. This occurs when the components of the Bloch functions are polynomials in $e^{\pm i\vc{k}\cdot\vc{a}_j}$, where $\vc{a}_j$ are the fundamental lattice vectors of the Bravais lattice. It is not known in general under which conditions a basis of Wannier functions with this property, called in the following compact Wannier functions, exists for a given band. However if one is interested just in a complete set of compact Wannier functions, which are not required to be linearly independent (a less restrictive condition than (\ref{eq:Wannier_normalization})), a complete answer exists~\cite{Read:2017}. In one dimension it is always possible to find compact Wannier functions in this generalized sense (called \qql compactly supported Wannier-type functions\qqr in Ref.~\cite{Read:2017}). In higher dimensions they exist only if all of the topological invariants characterizing $d\geq 2$ noninteracting topological phases are zero. One can only allow for nonzero $d=1$ topological invariants along each spatial direction, which characterize so-called \qql weak\qqr topological phases. See Ref.~\cite{Read:2017} for details.

For all the three 1D models (Creutz ladder, diamond chain and 1D dice lattice) analyzed here complete orthonormal sets (satisfying Eq.~(\ref{eq:Wannier_normalization})) of compact Wannier functions exists. In the case of the 2D dice lattice these can also be found for the lower and upper bands, but compact Wannier functions only in the generalized sense of Ref.~\cite{Read:2017} can apparently be found for the middle one, as shown in Appendix~\ref{app:compact_Wannier}. The polynomial Bloch functions that generate all of the compact Wannier function sets used in the following are provided in Appendix~\ref{app:compact_Wannier}.

Using orthonormal sets of compact Wannier functions is crucial for our purposes since they are used to expand the field operators 
\begin{equation}\label{eq:Wannier_expansion}
\hat{c}_{\vc{i}\alpha} = \sum_{n,\vc{j}} w_n(\vc{i}-\vc{j},\alpha)\hat{d}_{n\vc{j}}\,.
\end{equation}
The orthonormality of the Wannier functions ensures that the annihilation and creation operators in the Wannier function basis $\hat{d}_{n\vc{j}},\,\hat{d}_{n\vc{j}}^\dagger$ satify the (anti-)commutation relations for bosons (fermions). Using the orthonormality of the Wannier functions Eq.~(\ref{eq:Wannier_expansion}) can be inverted
\begin{equation}\label{eq:Wannier_expansion_inverse}
\hat{d}_{n\vc{j}}^\dagger = \sum_{\vc{i},\alpha}w_n(\vc{i}-\vc{j},\alpha)\hat{c}_{\vc{i}\alpha}^\dagger\,.
\end{equation}
The action of the conserved quantities on the operators $\hat{d}_{n\vc{j}},\,\hat{d}_{n\vc{j}}^\dagger$ turns out to be particularly simple. Consider for instance the Creutz ladder. By combining the gauge transformations
\begin{equation}\label{eq:G_Creutz_def}
\mathcal{\hat G}_{i} \hat{c}_{j\alpha} \mathcal{\hat G}^\dagger_{i}=
\begin{cases}
-\hat{c}_{j\alpha} & i=j,\, \\
\hat{c}_{j\alpha} & i\neq j\,.
\end{cases}
\end{equation}
with adjacent intertwining operators~(\ref{eq:U_creutz_1})-(\ref{eq:U_creutz_2})
one obtains the operators $\mathcal{\hat C}_i = \mathcal{\hat G}_{i}\mathcal{\hat U}_{i}\mathcal{\hat U}_{i+1}$ (see Fig.~\ref{fig:intertwining}), which commute with the full many-body Hamiltonian ($[\mathcal{\hat C}_i,\mathcal{\hat H}(\Phi)]=0$) for quite a general class of interaction terms, as previously discussed. In the specific case of the Creutz ladder the expansion in Eq.~(\ref{eq:Wannier_expansion_inverse})
reads (see Eq.~(\ref{eq:Wannierf_Creutz}) in the Appendix)
\begin{equation}
\hat{d}_{\pm,j}^\dagger = \frac{1}{2}\left(\hat{c}^\dagger_{j,1}+\hat{c}^\dagger_{j,2}\pm\hat{c}^\dagger_{j+1,1}\mp \hat{c}^\dagger_{j+1,2}\right)\,.
\end{equation}
It is easy to check from the above equation and the definition of the $\mathcal{\hat{C}}_i$ that
\begin{equation}\label{eq:cons_quant_action_d}
\mathcal{\hat{C}}_i\hat{d}_{\pm,j}^\dagger \mathcal{\hat{C}}_i^\dagger = \begin{cases}
-\hat{d}_{\pm,j}^\dagger & i = j\,, \\
\hat{d}_{\pm,j}^\dagger & i\neq j\,.
\end{cases}
\end{equation}
The last equation completely specifies the action of $\mathcal{\hat C}_i$ on the whole Hilbert space, therefore an equivalent form for these operators is
\begin{equation}\label{eq:parity_1}
\mathcal{\hat C}_j = \exp\Big(i\pi\sum_{n=\pm}\hat{d}_{nj}^\dagger\hat{d}_{nj}\Big)\,.
\end{equation}
In the spinful case one simply adds a summation over the spin degree of freedom
\begin{equation}\label{eq:parity_2}
\mathcal{\hat C}_j = \exp\Big(i\pi\sum_{n=\pm}\sum_{\sigma=\uparrow,\downarrow}\hat{d}_{nj\sigma}^\dagger\hat{d}_{nj\sigma}\Big)\,.
\end{equation}
The interpretation of Eq.~(\ref{eq:parity_1}) and~(\ref{eq:parity_2}) is clear. They keep track of the parity of the number of particles occupying the compact Wannier functions which have support on unit cells $j$ and $j+1$, in our convention. This is the parity of the occupation numbers summed over band and spin degrees of freedom, and possibly other internal degrees of freedom present in more general models.
If the occupation number is odd, that is if the system is in an eigenstate of $\mathcal{\hat C}_j$ with eigenvalue $-1$, there is an unpaired particle that may perform some dynamics in the spin or band space, but will always be localized in the group of compact Wannier functions labelled by the same unit cell index $j$. On the other hand, pairs of particles can be freely removed from or added on the same Wannier functions and in the presence of interactions they will most likely do so and will be found in states extended throughout the whole system. These extended states are responsible for the change of the ground state energy $E_0(\Phi)$ as the flux is varied. 


Consider then the 1D dice lattice of Fig.~\ref{fig:lattice_models}\textbf{c}. We are able to obtain conserved quantities only for the values $k_2 = 0,\pi$ for which the Hamiltonian is time-reversal invariant. In both cases $k_2 = 0,\pi$ the gauge transformation $\mathcal{\hat G}_i$ to be combined with the intertwining operators is the same as in Eq.~(\ref{eq:G_Creutz_def}), where the orbital index runs now over $\alpha=1,\dots,6$. The compact Wannier orbitals corresponding to the operators $\hat{d}_{n,j}$ are given implicitly in terms of polynomial Bloch functions in Appendix~\ref{app:compact_Wannier}, Eqs.~(\ref{eq:g_1_dice})-(\ref{eq:dice_norm}). 
Then the conserved quantity $\mathcal{\hat C}_i = \mathcal{\hat G}_{i}\mathcal{\hat U}_{i}\mathcal{\hat U}_{i+1}$ for $k_2 = 0$ can be written in the equivalent form
\begin{equation}\label{eq:Cj_Def_dice}
\mathcal{\hat C}_j = \exp\Big(i\pi\sum_{n=1,3,5}\hat{d}_{n,j}^\dagger\hat{d}_{n,j}+i\pi\sum_{n=2,4,6}\hat{d}_{n,j+1}^\dagger\hat{d}_{n,j+1}\Big)\,.
\end{equation}
The physical content of this last equation is the same as Eq.~(\ref{eq:parity_1}), that is a parity operator detecting unpaired particles in the occupation of a group of Wannier orbitals. Spin can be accounted for by taking two copies of Eq.~(\ref{eq:Cj_Def_dice}) as done in Eq.~(\ref{eq:parity_2}) for the Creutz ladder.
On the other hand the conserved quantities for $k_2 = \pi$ take exactly the same form as in Eq.~(\ref{eq:parity_1}) where the band index runs over $n = 1,\dots,6$.

The diamond chain is obtained from the 1D dice lattice by setting all the hoppings along vertical links to zero. These are the $t_1$ and $t_4$ hopping matrix elements in the notation of Appendix~\ref{app:2D_dice_Ham}. In fact these hoppings can be continuously decreased while preserving the perfect flatness of all bands. The conserved quantities of the 1D dice lattice for $k_2 = 0,\pi$ are preserved if $t_1 = t_4 = t'$ with $t'$ real but otherwise arbitrary. In this way one recovers the conserved quantities of the diamond chain~\cite{Doucot:2002} (see Fig.~\ref{fig:intertwining}) having the same form as in Eq.~(\ref{eq:parity_1}), where the band index runs over the three flat bands of the diamond chain ($n = 0,\pm$ in the notation of Appendix~\ref{app:diamond_chain_Ham} and~\ref{app:bands_diamond_chain}).

Using the commutation rules between field operators $\hat{d}_{nj}$ in the Wannier function basis and local conserved quantities $\mathcal{\hat C}_{j}$ in Eq.~(\ref{eq:cons_quant_action_d}), which are valid for all lattice models with suitable relabellings, it is easy to show that the single-particle propagator $\langle \hat{c}_{i\alpha}(t)\hat{c}_{j\beta}^\dagger(t')\rangle$ is in fact vanishing beyond a finite range~\cite{Kazymyrenko:2005}. For example for the Creutz ladder and diamond chain one finds that $\langle \hat{c}_{i\alpha}(t)\hat{c}_{j\beta}^\dagger(t')\rangle =0$ for $|i-j|\geq 2$, a result which has been confirmed in a recent DMRG study of the Creutz ladder~\cite{Mondaini:2018}. This is a consequence of the fact that the Wannier functions are compact. On the other hand if the Wannier functions were just exponentially decaying then the single-particle propagator would be exponentially decaying and thus short-ranged as well. 
A short-ranged propagator indicates precisely that single-particle excitations are localized in a quantum many-body system. On the other hand we know for the same lattice models that the ground state is conductive (see Section ~\ref{sec:boundary_condition} and~\ref{sec:numerics}). Therefore we conclude that charge transport is solely a result of the motion of $2n$-body bound states, in particular Cooper pairs ($n =1$) as shown by the general arguments of Ref.~\cite{Tovmasyan:2016}. These pairs are preformed since they are present even above the critical temperature to a superfluid state. Indeed, our argument shows that the single-particle propagator is short-ranged at any temperature. 

Note that the locality of the conserved quantities is essential to prove that the single-particle propagator is short-ranged. The global intertwining operators of Appendic~\ref{app:global_U} cannot be used to constrain the single-particle propagator, precisely because of their global nature.

Instead of Eq.~(\ref{eq:cons_quant_action_d}) one can consider the family of unitary operators $\mathcal{\hat{C}}_i(\phi)$ parametrized by $\phi$ and defined by their action on the field operators in the Wannier basis
\begin{equation}\label{eq:cons_quant_general}
\mathcal{\hat{C}}_i(\phi)\hat{d}_{\pm,j}^\dagger \mathcal{\hat{C}}_i^\dagger(\phi) = \begin{cases}
e^{i\phi}\hat{d}_{\pm,j}^\dagger & i = j\,, \\
\hat{d}_{\pm,j}^\dagger & i\neq j\,.
\end{cases}
\end{equation}
By construction these commute with the Creutz ladder Hamiltonian $\mathcal{\hat H}_0$. Obviously the class of interaction terms that commutes with the operators $\mathcal{\hat{C}}_i(\phi)$ for arbitrary $\phi$ is contained in the one commuting with $\mathcal{\hat{C}}_i(\phi = \pi) = \mathcal{\hat{C}}_i$ only. In some cases they actually coincide. This is the case of spinless fermions on the Creutz ladder studied in Ref.~\cite{Junemann:2017}. Indeed one finds that the operators $\mathcal{\hat C}_i(\phi)$ perform canonical transformations of the field operators $\hat{c}_{i,\alpha}$ which mix orbitals within the same unit cell only, therefore the operators $\hat{n}_j =\hat{c}_{j,1}^\dagger\hat{c}_{j,1}+\hat{c}_{j,2}^\dagger\hat{c}_{j,2}$ are invariant for all $j$. Thus the most general quartic interaction term preserving the conserved quantities for arbitrary $\phi$ has the form $\mathcal{\hat H}_{\rm int} = \sum_{ij}V_{ij}\hat{n}_i\hat{n}_j$. 
On the other hand it is easy to check that any quartic interaction term commuting with the operators $\mathcal{\hat{C}}_i$ has exactly the same form. 
For example, using the fermionic commutation relations, the nearest-neighbour interaction term considered in Ref.~\cite{Junemann:2017} can be written in the form 
\begin{equation}
\mathcal{\hat H}_{\rm int} = V\sum_{j}\hat{c}_{j,1}^\dagger\hat{c}_{j,1}\hat{c}_{j,2}^\dagger\hat{c}_{j,2} = \frac{V}{2}\sum_{j}(\hat{n}^2_j-\hat{n}_j)\,,
\end{equation}
which is manifestly invariant under the action of $\mathcal{\hat C}_i(\phi)$ for arbitrary $\phi$. The good quantum numbers are in this case the total particle number in the Wannier orbitals labelled by the same $j$ (the eigenvalue of $\hat{d}_{+,j}^\dagger\hat{d}_{+,j} + \hat{d}_{-,j}^\dagger\hat{d}_{-,j}$), not just its parity as in the case of spin-$1/2$ fermions. This means that particles cannot move around in the lattice and transport is suppressed in any form. Similar considerations apply also to the 1D dice lattice and were made in the case of the diamond chain already in Ref.~\cite{Doucot:2002}. The on-site Hubbard interaction is special since both in the case of bosons and spin-$1/2$ fermions it is invariant under $\mathcal{\hat C}_i(\phi)$ strictly only for $\phi = \pi$, and thus allows pair transport. 

We conclude this section by observing that in all the models considered here the conserved quantities have the same form: parity operators associated to the occupation of compact Wannier orbitals. Therefore orthogonal sets of compact Wannier functions seem to play an essential role in the construction of these conserved quantities. In the next section we analyze in more detail some of the consequences of the presence of these conserved quantities. 

\subsection{Projection on the flat band}
\label{sec:projected_conserved_quantities}

The presence of an extensive number of mutually commuting local integrals of motion is useful for a number of purposes. As an example, here we use the results of the previous section to explain some findings regarding the fermionic Creutz ladder with Hubbard interaction presented in our previous work~\cite{Tovmasyan:2016}.

The operators $\mathcal{\hat C}_j$ do not commute only with the full many-body Hamiltonian $\mathcal{\hat H}= \mathcal{\hat H}_0+\mathcal{\hat H}_{\rm int}$, but also with the noninteracting part $\mathcal{\hat H}_0$ and the interaction term $\mathcal{\hat H}_{\rm int}$ separately. Therefore these are conserved quantities also of the Hamiltonian projected on the flat band many-body subspace. The technique of projecting a Hamiltonian on the flat band of the noninteracting term is useful when the interaction strength is smaller than or comparable to the band gap separating the flat band from other bands. This technique has been routinely used to explore the properties of flat band models with interactions, both in the case of bosons~\cite{Huber:2010,Tovmasyan:2013,Takayoshi:2013} and fermions~\cite{Vidal:2001,Tovmasyan:2016}. The general formalism is presented in detail in Ref.~\cite{Tovmasyan:2016} where it is viewed as the first order step in a perturbative expansion in the ratio between interaction strength and band gap $|U|/E_{\rm g}$. 
In the case of spinful fermions the Hubbard interaction projected on the lowest band of the Creutz ladder has been calculated in Ref.~\cite{Tovmasyan:2016} and this result is summarized here.

The field operators projected on the lowest band $n = -$ of the Creutz ladder are defined by truncating the sum over bands in Eq.~(\ref{eq:Wannier_expansion})
\begin{equation}
\bar{c}_{i\alpha\sigma} = \sum_{j} w_-(i-j,\alpha)\hat{d}_{-,j\sigma}\,.
\end{equation}
We use a bar ($\bar{c}$) instead of a hat ($\hat{c}$) to denote operators projected on the lowest band, as in Ref.~\cite{Tovmasyan:2016}. The projected Hubbard interaction is then 
\begin{equation}\label{eq:projected_Hubbard}
\overline{\mathcal{H}}_{\rm int} = U\sum_{i,\alpha}\bar{c}_{i\alpha\uparrow}^\dagger\bar{c}_{i\alpha\uparrow}\bar{c}_{i\alpha\downarrow}^\dagger\bar{c}_{i\alpha\downarrow}\,.
\end{equation}
Contrary to Ref.~\cite{Tovmasyan:2016}, we use here the convention that the interaction parameter $U$ can be both positive (repulsive case) or negative (attractive case).
In order to expand the projected interaction~(\ref{eq:projected_Hubbard}) in terms of the operators in the Wannier basis it is convenient to introduce two families of (pseudo-)spin operators. One family of operators encodes the spin degree of freedom of localized single particles in the Wannier orbitals of the lowest band
\begin{gather}
\hat{S}_j^{+} = \big(\hat{S}_j^{-}\big)^\dagger = \hat{d}_{-,j\uparrow}^\dagger\hat{d}_{-,j\downarrow}\,,\\
\hat{S}_j^{z} = \frac{1}{2}\big(\hat{d}_{-,j\uparrow}^\dagger\hat{d}_{-,j\uparrow}-\hat{d}_{-,j\downarrow}^\dagger\hat{d}_{-,j\downarrow}\big)\,,\\
\hat{S}_j^{x} = \frac{1}{2}\big(\hat{S}_j^{+}+\hat{S}_j^{-}\big)\,,\qquad\hat{S}_j^{y} = \frac{1}{2i}\big(\hat{S}_j^{+}-\hat{S}_j^{-}\big)\,.
\end{gather}
These operators satisfy the usual $SU(2)$ commutation relations. The other family of (pseudo-)spin operators encodes the presence or absence of a pair of particles in the same orbitals
\begin{gather}
\hat{T}_j^{+} = \big(\hat{T}_j^{-}\big)^\dagger = \hat{d}_{-,j\uparrow}^\dagger\hat{d}_{-,j\downarrow}^\dagger\,,\\
\hat{T}_j^{z} = \frac{1}{2}\big(\hat{d}_{-,j\uparrow}^\dagger\hat{d}_{-,j\uparrow}+\hat{d}_{-,j\downarrow}^\dagger\hat{d}_{-,j\downarrow}-1\big)\,,\\
\hat{T}_j^{x} = \frac{1}{2}\big(\hat{T}_j^{+}+\hat{T}_j^{-}\big)\,,\qquad\hat{T}_j^{y} = \frac{1}{2i}\big(\hat{T}_j^{+}-\hat{T}_j^{-}\big)\,.
\end{gather}
They satisfy the same commutation relations as the $\hat{S}_j^{(x,y,z)}$, moreover the two set of spin operators commute with each other $[\hat{S}_j^\alpha,\hat{T}_l^\beta] = 0$.
The two families of operators are related by a particle-hole transformation on the $\downarrow$-spin ($\hat{d}_{-,j\downarrow} \rightarrow \hat{d}_{-,j\downarrow}^\dagger$). In terms of these operators the projected Hamiltonian is
\begin{equation}\label{eq:project_Ham_spin_form}
\begin{split}
\overline{\mathcal{H}}_{\rm int} &= \frac{U}{4}\sum_{j}\hat{\vc{T}}_j\cdot\hat{\vc{T}}_{j+1} -\frac{U}{4}\sum_j \hat{\vc{S}}_j\cdot\hat{\vc{S}}_{j+1} \\
&+\frac{U}{4}\sum_{j}\left(\hat{T}_j^+\hat{T}_j^-+\hat{T}^z_j+\frac{1}{4}\right)
\,,
\end{split}
\end{equation}
where $\hat{\vc{T}}_j = (\hat{T}_j^x,\hat{T}_j^y,\hat{T}_j^z)^T$ and $\hat{\vc{S}}_j = (\hat{S}_j^x,\hat{S}_j^y,\hat{S}_j^z)^T$.
The fact that the projected Hamiltonian can be expressed entirely in terms of the two families of pseudospin operators is a consequence of the local conserved quantities of the Creutz-Hubbard model with spin, Eq.~(\ref{eq:parity_2}). Indeed it is easy to prove that $\overline{\mathcal{H}}_{\rm int}$ commutes with 
\begin{equation}\label{eq:proj_conserved_quantity}
\mathcal{\overline{C}}_j = \exp\Big(i\pi\sum_{\sigma=\uparrow,\downarrow}\hat{d}_{-,j\sigma}^\dagger\hat{d}_{-,j\sigma}\Big)\,,
\end{equation}
the projected form of Eq.~(\ref{eq:parity_2}). Viceversa any Hamiltonian commuting with all the $\mathcal{\overline{C}}_j$ can be expressed only in terms of the two families of spin operators. Indeed, take an Hamiltonian which is a linear combination of terms of the form $\hat{d}_{j_{1}\uparrow}^\dagger\hat{d}_{j_{2}\downarrow}^\dagger\hat{d}_{j_{3}\downarrow}\hat{d}_{j_{4}\uparrow}$ such as Eq.~(\ref{eq:projected_Hubbard}). By requiring that each of them commutes with all the operators $\mathcal{\overline{C}}_j$ one obtains the constrain that the indices $j_1,j_2,j_3,j_4$ are two by two equal. From this it follows easily that each term in the linear combination can be rewritten in terms of the pseudospin operators. For example the case $j_1 = j_2$ and $j_3 = j_4$ corresponds to a pair-hopping term $\hat{T}_{j_1}^+\hat{T}^-_{j_3}$.

It may seem from Eq.~(\ref{eq:project_Ham_spin_form}) that the projected Hamiltonian is equivalent to two uncoupled copies of identical spin chains with opposite sign of the couplings. However, the Pauli exclusion principle imposes some constrains on the Hilbert space on which the projected Hamiltonian acts. Denote by $S_j$ and $T_j$ the total angular momentum of the pseudospin operators defined as usual by the eigenvalues of $\hat{\vc{S}}_{j}^2$ and $\hat{\vc{T}}_{j}^2$, respectively. Then the constrains are
\begin{gather}
\mathcal{\overline{C}}_j\ket{\psi} = +\ket{\psi} \Rightarrow S_j= 0\,,\,\,T_j = 1/2 \label{eq:constrain_1}\\
\mathcal{\overline{C}}_j\ket{\psi} = -\ket{\psi} \Rightarrow S_j= 1/2\,,\,\,T_j = 0 \,, \label{eq:constrain_2}
\end{gather}
where $\ket{\psi}$ is an arbitrary state in the allowed Hilbert space. These constrains express the fact that pairs and unpaired particles cannot reside simultaneously in the same spinful Wannier orbital. In particular, an unpaired particle at Wannier orbital $j$ prevents all pairs on the left side to tunnel on the right side since only nearest neighbor pair-hopping terms (of the form $\hat{T}_{j}^+\hat{T}_{j+1}^-+\text{H.c.}$) are present in Eq.~(\ref{eq:project_Ham_spin_form}). At the level of the projected Hamiltonian transport is completely suppressed by a single localized particle.

However, transport is allowed even in the presence of unpaired particles if interband transitions are taken into account. This amounts to take higher orders in the expansion in $|U|/E_{\rm g}$. The second order in the expansion has been calculated in Ref.~\cite{Tovmasyan:2016}, where it is noted that even at this order terms violating the conserved quantities in Eq.~(\ref{eq:proj_conserved_quantity}) are absent. This means that even at the second order in perturbation theory, the low energy effective Hamiltonian can be expressed solely in terms of the (pseudo-)spin operators and the motion of unpaired particles is completely suppressed. With the result of the present work, we now understand that this is not accidental, but occurs at any order of perturbation theory due to the exact symmetries of the full many-body Hamiltonian~(\ref{eq:parity_2}). At second order terms of the form $\hat{T}_{j-1}^+\hat{T}^-_{j+1}+\text{H.c.}$ do appear (see Eq.~(44)-(45) in Ref.~\cite{Tovmasyan:2016}) which allow a pair of particles to jump over a localized particle in Wannier orbital $j$. Consistently in the presence of spin imbalance, we find that $|E_0(\Phi)-E_0(0)| \propto U^2/E_{\rm g}$ for small $U$, see Fig.~\ref{fig:Creutz_num_res_2}.

We observe in passing that a similar situation occurred in Ref.~\cite{Coester:2013}, where local $\mathbb{Z}_2$ symmetries of the low energy effective Hamiltonian of the quantum Ising model on the diamond chain were found, but were not associated to conserved quantities of the full Hamiltonian.

A final interesting observation is that the projected Hamiltonian $\overline{\mathcal{H}}_{\rm int}$ is completely integrable in the Bethe ansatz sense. Indeed one can fix a subspace of the Hilbert space by specifying all the eigenvalues $C_j= \pm1$ of $\mathcal{\overline{C}}_j$. In this subspace the Hamiltonian breaks down in independent $XXX$ spin chains~\cite{Korepin_Book} with both open or periodic boundary conditions for the degrees of freedom of spin ($C_j = -1$) and pair propagation ($C_j= +1$). The $XXX$ spin chain with both open and periodic boundary conditions can be solved by Bethe ansatz~\cite{Wang_book}. In view of the previous discussion a natural question for future studies is whether integrability is an approximate low-energy property only or a feature of the full fermionic Creutz-Hubbard mode and possibly other models.

\section{Discussion and conclusion}
\label{sec:disc_concl}

In this work we address the question of the nature of the charge carriers in a flat Bloch band of a lattice Hamiltonian in the presence of interparticle interactions. The main result is that in some one dimensional lattice models single-particle transport is completely suppressed since the single-particle propagator is short-ranged at any temperature, in fact strictly vanishing beyond a finite range. This is a consequence of the extensive number of independent local conserved quantities constructed in this work. This result, combined with our previous finding that the superfluid weight is nonzero in the same models in the presence of an attractive Hubbard interaction~\cite{Peotta:2015,Julku:2016,Tovmasyan:2016,Liang:2017a,Liang:2017b}, imply that pairs of particles are necessarily present and are the only charge carriers. Moreover, we find that pair transport manifests as the period halving of the Aharonov-Bohm effect in a ring geometry. Indeed, we consistently observe in our numerical simulations that the ground state energy is $\Phi_0/2$-periodic as a function of the magnetic flux threading the ring. Interestingly, the local intertwining operators of Sec.~\ref{sec:graph_automorphism}, which show that the ground state energy is exactly $\Phi_0/2$-periodic, are also the building blocks of the local conserved quantities.

Our exact analytical arguments are rather powerful and have general applicability. The intertwining operators for 1D lattice Hamiltonians show that the entire many-body spectrum, not just the ground state energy, is $\Phi_0/2$-periodic. Therefore signatures of this peculiar periodicity should be observable even at high temperature. The intertwining operators are canonical transformations, in the specific permutations of field operators modulo a gauge transformation, as a consequence our argument applies to bosons, fermions and even hardcore bosons and spin systems~\cite{Coester:2013}, emphasizing that this phenomenon is purely a result of the band flatness and independent of the specific details of the lattice Hamiltonian and particle statistics.

The fact that the ground state energy and the associated persistent current have a half-flux quantum period was anticipated in Ref.~\cite{Doucot:2002} in the case of the diamond chain and verified in the context of Josephson junction arrays both in the clean ~\cite{Protopopov:2004} and disordered cases~\cite{Protopopov:2006} by explicitly calculating the persistent current. The anomalous periodicity of the current in Josephson junction arrays was subsequently measured in Refs.~\cite{Pop:2008,Gladchenko:2008}. Our results show that the $\Phi_0/2$-periodicity of the current is robust and independent of any approximation and applies to more general systems than Josephson junction arrays, on which no experiments have been performed yet.


We have shown in the case of the 1D dice lattice that local conserved quantities may or may not exists depending on the value of the hopping matrix elements even if the bands remain flat. The periodicity of the ground state energy can serve as a simple effective tool to probe the existence of such conserved quantities, in particular in the case of spin imbalance for spin-$1/2$ fermions as shown in Figs.~\ref{fig:1D_dice}-\ref{fig:2D_dice}-\ref{fig:fitting}. An interesting direction for future work is to analytically show that the ground state energy is $\Phi_0/2$-periodic in the spin balanced case, as pointed by our numerical results, and investigate the nature of the excitations responsible for the breaking of this periodicity in the case of spin imbalance. Figs.~\ref{fig:2D_dice}-\ref{fig:fitting} show that the lifting of $\Phi_0/2$-periodicity is due to interband mixing in the case of spin imbalance, rather than being an effect intrinsic to the partially filled flat band. Thus it might be possible to provide rigorous analytical arguments at the level of the projected Hamiltonian as done in Sec.~\ref{sec:projected_conserved_quantities}. For example it might be possible to prove that the ground state energy is $\Phi_0/2$-periodic up to a correction of order $(U/E_{\rm g})^2$ or higher. Indeed in all cases the local conserved quantities commute with the noninteracting part and the interaction term of the Hamiltonian separately, which means that they survive the projection of the interaction term onto the flat band many-body subspace.

As mentioned before, the local conserved quantities had already been found  for the diamond chain in Ref.~\cite{Doucot:2002}, while the fact that interactions can lead to the delocalization of two-body bound states was pointed out in Ref.~\cite{Vidal:2000}. The novel point of the present work is that these are not peculiarities of a special lattice model such as the diamond chain, but rather generic features of lattices with flat bands. Of course we are not able to provide completely general arguments, but our results point out that this should be possible to some extent.

Our approach has two main advantages. First it connects the existence of local conserved quantities to an exact observable property, the half-flux quantum periodicity of the persistent current. Second it emphasizes the connection between these symmetries and canonical transformations built from graph automorphisms. The connection between graph theory and flat band physics has a long history~\cite{Mielke:1991,Mielke:1991b,Mielke:1992,Mielke:1993}
and recent results~\cite{Drescher:2017,Roentgen:2018} indicate that much remains to be done in this direction. Moreover it is clear that our results apply regardless of particle statistics. Curiously, much of the work that followed the seminal work by Dou\c{c}ot and Vidal~\cite{Doucot:2002} has focused on bosons, more specifically on Josephson junction arrays~\cite{Ioffe:2002,Protopopov:2004,Protopopov:2006,Rizzi:2006,Rizzi:2006a,Pop:2008,Gladchenko:2008}, also in view of possible applications to topological quantum computing~\cite{Doucot:2012}.
There are some theoretical works on the diamond chain with fermions~\cite{Gulacsi:2007,Movilla:2011,Lopes:2011,Rojas:2012,Hyrkaes:2013,Lopes:2014,Kobayashi:2016}, however the question of the existence of the local conserved quantities has not been explicitly addressed in any of these works. The only instance where local $\mathbb{Z}_2$ symmetries have been discussed in the case of fermions is Ref.~\cite{Kazymyrenko:2005}, which considers the different case of quantum wire networks, rather than lattice models. 

In our opinion the fermionic case deserves more attention than what has been paid to it so far, indeed, \qql assuming that the local $\mathbb{Z}_2$ symmetry
cannot be spontaneously broken, \dots such a system is
never a Landau-Fermi liquid, since the single-electron propagator is short ranged in space\qqr~\cite{Kazymyrenko:2005}. This means that lattices with flat bands can serve as a paradigm for the study of non-Fermi liquid behavior. In particular in a flat band transport seems to be dominated by two-body bound states even at high temperature. This is an exact result in the one-dimensional lattice models for which we are able to provide explicitly the intertwining operators and the local conserved quantities, but our numerical results suggest that this might be approximately true also for models in higher dimensionality, such as the 2D dice lattice. This suggests an analogy with the enigmatic pseudogap phase observed in underdoped cuprates above the transition temperature~\cite{Rice:2012,Stewart:2017}, for which an explanation in terms of preformed pairs has been put forward~\cite{Randeria:1998}. Moreover, recent experiments~\cite{Cao:2018} on twisted bilayer graphene have found an unconventional superconducting state precisely in correspondence of the twisting angle at which the moir\'e bands are quasi-flat~\cite{Bistritzer:2011}. This superconductive state emerges from a doped Mott insulator and therefore is intriguingly similar to the one found in cuprates. We advocate that the flat band limit can be a useful staring point for understanding these unconventional superconducting states. One reason is that in the flat band limit exact analytical results can be obtained, as shown here. These exact results can be used to benchmark the approximations necessary to tackle more general models.

There are several avenues where our results could be put to experimental test. In recent years there have been several realizations of flat bands models using optical lattices~\cite{Jo:2012,Taie:2015,Ozawa:2017,Taie:2017} and more recently the Creutz ladder has been proposed as a workhorse for the study of topological effects in ultracold gases and its implementation seems to be within reach with current experimental tools~\cite{Junemann:2017}. The diamond chain has been recently implemented with photonic lattices~\cite{Mukherjee:2018}, while excited orbital angular momentum states of ultracold atoms have been proposed as a new venue for implementing the same model~\cite{Pelegri:2018}. We note also a recent theoretical work~\cite{Cartwright:2018} where the phase diagram of the diamond chain with a Bose-Hubbard interaction term has been studied in detail and possible strategies for its implementation have been described.

Another important achievement has been the implementation of the hexagonal lattice~\cite{Tarruell:2012}, from which a topologically nontrivial band structure is obtained by shaking the optical lattice~\cite{Jotzu:2014,Flaeschner:2016}. The lowest band of the hexagonal lattice can be made almost flat by tuning the strength of the shaking~\cite{Neupert:2011,Liang:2017a}. 
The 2D dice lattice, closely related to the hexagonal lattice, has not been implemented yet, but a concrete proposal in this sense has been put forward recently~\cite{Gunnar:2018}.
Therefore ultracold gases are a promising platform for the study of flat band physics and our results show that interesting effects could be observable even at high temperature. Indeed a current challenge for experiments with ultracold atomic gases in optical lattices is the need to achieve lower temperatures to uncover interesting phases of matter, for example quantum magnetism and unconventional superfluid states. Our work points out that signatures of an exotic normal state characterized by the presence preformed pairs should be observable at any temperature in lattices with flat bands. In this sense it would be interesting to study how robust are the effects due to preformed pairs in the presence of perturbations that slightly break the perfect flatness of the bands and their characteristic temperature scale. 
 
In the solid state context we envisage the use of quantum dot arrays, or artificial nanostructures~\cite{Drost:2017,Slot:2017} to observe the phenomenon of period halving of the Aharonov-Bohm effect in non-superconducting electronic systems, in analogy to what has been done in the case of Josephson junctions arrays. 
Beyond ultracold gases and electronic systems, there are nowadays a number of available experimental platforms that can be used to explore flat band physics, see Ref.~\cite{Leykam:2018} for a recent review.
We hope that our work will stimulate further theoretical and experimental investigations in this direction.

\acknowledgements
\noindent This work was supported by the Academy of Finland through its Centers of Excellence Programme (2012-2017) and under Project Nos. 303351, 307419, and 284621 and by the European Research Council (ERC-2013-AdG-340748-CODE). S.P. acknowledges funding from the European Union's Horizon 2020 research and innovation programme under the Marie Sk\l{}odowska-Curie grant agreement No. 702281 (FLATOPS). M.T. and S.D.H. acknowledge financial support from the Swiss National Science Foundation and the NCCR QSIT. Computing resources were provided by the Euler high-performance computer at the Swiss National Supercomputing Centre, CSC - the Finnish IT Centre for Science and the Triton cluster at Aalto University.

\textit{Note} --- After the submission of this work for publication, a related independent work appeared in arXiv, Ref.~\cite{Mondaini:2018}. In this latter reference an in depth DMRG analysis of the Creutz ladder has been performed, in particular the finite size scaling of the energy vs. flux $E_0(\Phi)$ and of the Drude weight, as in Fig.~\ref{fig:Drude_Creutz} of this work. The numerical results of Ref.~\cite{Mondaini:2018} provide an independent verification of the validity of Eqs.~(\ref{eq:exact_Ds})-(\ref{eq:QGT}) and confirm that the single-particle propagator is short-ranged, which is a consequence of the existence of the conserved quantities found in this work for the Creutz ladder, Eq.~(\ref{eq:parity_2}).

\appendix

\section{Hamiltonians of lattice models}
\label{app:lattice_Ham}

\subsection{Creutz ladder}
\label{app:def_Creutz_ladder}

The Creutz ladder is shown schematically in Fig.~\ref{fig:lattice_models}\textbf{a}. 
We provide the hopping matrix of a slightly more general model where the hoppings along the horizontal direction ($\alpha =1 \leftrightarrow \alpha = 1$, $\alpha = 2 \leftrightarrow \alpha = 2$) can have a different magnitude from the ones in the diagonal direction ($\alpha =1\leftrightarrow \alpha =2$). The two complex hopping matrix elements are called $t_1,t_2$. This more general model where the bands are nonflat for $t_1\neq \pm t_2$ is used in Fig.~\ref{fig:Creutz_num_res_1} in the bottom panel.
The corresponding tight-binding Hamiltonian is
\begin{equation}
\label{eq:Creutz_def1}
\begin{split}
\mathcal{\hat H}_{\rm Creutz} &= \sum_{j}\Big[t_1\Big(\hat{c}^\dagger_{j+1,1}\hat{c}_{j,1}-\hat{c}^\dagger_{j+1,2}\hat{c}_{j,2}\Big) \\&\qquad +t_2\Big(\hat{c}^\dagger_{j+1,1}\hat{c}_{j,2}-\hat{c}^\dagger_{j+1,2}\hat{c}_{j,1}\Big)+\mathrm{H.c.}\Big]\,.
\end{split}
\end{equation}
The matrix $K(\vc{j}) = K(j)$ is nonzero only for $j = \pm 1$ and it reads
\begin{equation}
\label{eq:Creutz_def2}
K(1) = [K(-1)]^\dagger = \begin{pmatrix}
t_1 & t_2 \\
-t_2 & -t_1
\end{pmatrix}\,.
\end{equation} 
The lattice spacing is denoted by $a$, as a consequence for a lattice of length $L = N_{\rm c}a$ with periodic Born-von K\'arm\'an boundary conditions, the wavevectors are quantized in multiples of $2\pi/L$, that is $k = \frac{2\pi m}{L}$ with $m \in 0,\dots,N_{\rm c}-1$. 
Then from the definition of the Fourier transform of the hopping matrix one has
\begin{equation}
\label{eq:Creutz_def3}
\begin{split}
&\widetilde{K}(\vc{k}) = \widetilde{K}(k) = \\ &=\begin{pmatrix}
t_1 e^{-ika} + t_1^* e^{ika}& t_2e^{-ika}-t_2^*e^{ika} \\
t^*_2e^{ika}-t_2e^{-ika} & -t_1 e^{-ika} - t_1^* e^{ika}
\end{pmatrix} \\
&= (t_1 e^{-ika} + t_1^* e^{ika})\sigma_z +i (t_2e^{-ika}-t^*_2e^{ika})\sigma_y
\,,\end{split}
\end{equation}
where $\sigma_{x,y,z}$ are the Pauli matrices.

\subsection{Two dimensional dice lattice and dimensional reduction} \label{app:2D_dice_Ham}

The dice lattice is a tripartite lattice. There are two types of sites: the \textit{hub sites} have coordination number 6 (hexagons in Fig.~\ref{fig:lattice_models}\textbf{b}-\textbf{c}) and the \textit{rim sites} have coordination
number 3 (triangles in Fig.~\ref{fig:lattice_models}\textbf{b}-\textbf{c}). The rim sites
can be divided into two sublattices. In total there are three sublattices and all links
connect sites from two different sublattices.

If all the hopping matrix elements are equal the underlying Bravais lattice is triangular with fundamental vectors $\vc{v}_1 = (\sqrt{3}a,0)^T$, $\vc{v}_2 = (\sqrt{3}a/2,3a/2)^T$. Here $a$ is the length of the edge of an elementary rhombus in the lattice. The vectors connecting a hub sites to the rim sites are $\vc{b}_1 = (0,a)^T$, $\vc{b}_2 = (\sqrt{3}a/2,a/2)^T$ and $\vc{b}_2 = (\sqrt{3}a/2,-a/2)^T$ and their opposite $-\vc{b}_{i=1,2,3}$.

With the signs of the hopping matrix elements chosen as in Fig.~\ref{fig:lattice_models}\textbf{b}-\textbf{c} the flux through each elementary rhombus is half of a flux quantum. The area of the unit cell is three times the area of an elementary rhombus, which means that the flux through a unit cell is also half of a flux quantum (flux is defined modulo $2\pi$ in a lattice). Therefore the area of the smallest magnetic unit cell is twice as large as the unit cell of the triangular lattice. The magnetic unit cell used in this work is shown in Fig.~\ref{fig:lattice_models}\textbf{b}-\textbf{c} and comprises six lattices sites, four rim sites $\alpha = 1,2,5,6$ and the two hub sites $\alpha = 3,4$. The underlying Bravais lattice is rectangular and the fundamental vectors are $\vc{a}_1 = \vc{v}_1$ and $\vc{a}_2 = 2\vc{v}_2-\vc{v}_1 = (0,3a)^T$ as shown in Fig.~\ref{fig:lattice_models}\textbf{b}.

In Fig.~\ref{fig:lattice_models}\textbf{b}-\textbf{c} all the hopping matrix elements have the same magnitude $t$ up to a phase, as in indicated by the colour of the links. In fact one can allow for different amplitudes along the three distinct directions $\vc{b}_{i=1,2,3}$ of the links, while preserving the perfect flatness of the bands. Call $t_i$ the hopping amplitude along the links parallel to $\vc{b}_i$. On the vertical links (parallel to $\vc{b}_1$) we also alternate between hopping matrix elements with amplitude $t_1$ and $t_4$. One could even consider an even more general model with perfectly flat spectrum with alternating hopping matrix elements also along the diagonal directions $\vc{b}_{2,3}$. This is not done here since the magnetic unit cell is larger in this case.
The Fourier transform of the hopping matrix of the more general 2D dice lattice is
\begin{widetext}
\begin{equation}\label{eq:2D_dice_def}
\widetilde{K}(k_1,k_2) = 
\begin{pmatrix}
0 & 0 & t^*_3+t_2e^{-ik_1} & -t^*_1 & 0 & 0 \\
0 & 0 & t^*_2-t_3e^{-ik_1} & t_4e^{-ik_2} & 0 & 0 \\
t_3+t^*_2e^{ik_1} & t_2-t^*_3e^{ik_1} & \varepsilon_{\rm h} & 0 & t_4e^{-ik_2} & t^*_1 \\
-t_1 & t_4^*e^{ik_2} & 0 & \varepsilon_{\rm h} & -t^*_2+t_3e^{-ik_1} & t^*_3+t_2e^{-ik_1} \\ 
0 & 0 & t_4^*e^{ik_2} & -t_2+t_3^*e^{ik_1} & 0 & 0 \\
0 & 0 & t_1 & t_3+t^*_2e^{ik_1} & 0 & 0
\end{pmatrix}
\end{equation}
\end{widetext}
where $\vc{k}\cdot\vc{a}_1 = k_2$, $\vc{k}\cdot\vc{a}_2 = k_1$ and $k_1,k_2\in [-\pi,\pi]$. We also allow for different onsite energy for the rim ($\varepsilon_{\rm r} = 0$) and hub ($\varepsilon_{\rm h}$) sites. Even if $\varepsilon_{\rm h} \neq 0$ all the bands are still perfectly flat. In Fig.~\ref{fig:lattice_models}\textbf{b} we show the 2D dice lattice with $t_1 = t_2 = t_3 = t_4 = t$, the only case considered in our numerical simulations.

The 1D dice lattice is obtained by dimensional reduction of the 2D dice lattice. The hopping matrix of the 1D dice lattice is obtained by inverting the Fourier transform only with respect to $k_1$, while $k_2$ is in this case a continuous parameter which gives a family of 1D lattices with hopping matrix
\begin{equation}\label{eq:partial_FT}
K_{k_2}(i-j) = \frac{1}{2\pi} \int_{-\pi}^{\pi}dk_{1}\,\widetilde{K}(k_1,k_2)e^{ik_1(i-j)}\,.
\end{equation}
In Fig.~\ref{fig:lattice_models}\textbf{c} we show the 1D dice lattice for $t_1=t_2=t_3=t_4 =t$ and arbitrary $k_2$, the case on which we concentrate in this work.
Of particular interest are the values $k_2 = 0$ and $k_2 = \pi$, for which the Hamiltonian is time-reversal symmetric. The 1D dice lattice for $k_2 = 0$ corresponds to the choice of periodic boundary condition along the transverse ($\vc{a}_2$) direction, while $k_2 = \pi$ to antiperiodic boundary conditions along the same direction.

\subsection{Diamond chain}
\label{app:diamond_chain_Ham}

The diamond chain~\cite{Doucot:2002} can also be obtained from the 2D dice lattice by setting $t_1=t_4=0$ and eliminating sites $\alpha = 4,5,6$. Then the 2D lattice decouples in parallel 1D chains all identical to the diamond chain. This amounts to retaining only the minor relative to the first three columns/rows of the matrix in Eq.~(\ref{eq:2D_dice_def}), that is
\begin{equation}\label{eq:diamond_T_def}
\widetilde{K}(k) = \begin{pmatrix}
0 & 0 & t_3^*+t_2e^{-ik a} \\
0 & 0 & t_2^*-t_3e^{-ik a} \\
t_3+t_2^*e^{ik a} & t_2-t_3^*e^{ik a} & \varepsilon_{\rm h}
\end{pmatrix}\,.
\end{equation}
In the original diamond chain discussed in Ref.~\cite{Doucot:2002} the hopping matrix elements are all equal which amounts to setting $t_2=t_3=t$. The band structure is composed of perfectly flat bands for arbitrary values of the complex parameters $t_2,\,t_3$ and of the on-site energy $\varepsilon_{\rm h}$ of the hub sites. In Fig.~\ref{fig:lattice_models}\textbf{d} the case $t_2 = t_3 = t$ is represented.

\section{Compactly localized Wannier functions}
\label{app:compact_Wannier}

In this Appendix the compact Wannier functions for the lattice models considered in the present work are provided. In fact we provide the polynomial Bloch functions which are related to the compact Wannier functions by a Fourier transform~(\ref{eq:Wannier_func_def}). In the basis of compact Wannier functions the conserved quantities have a particularly simple interpretation as parity operators, as explained in Sec.~\ref{sec:interpretation}. 

\subsection{Creutz ladder}

The band structure of the Creutz ladder defined by the Hamiltonian in Eqs.~(\ref{eq:Creutz_def1})-(\ref{eq:Creutz_def3}) is composed of perfectly flat bands for $t_1 = \pm t_2$. Let us take the case $t_1 = t_2 = t$ where $t$ is real and positive. The energy eigenvalues for the two flat bands are then $\varepsilon_{\pm} = \pm 2t$ and the corresponding Bloch functions are
\begin{equation}
g_{\pm}(k) = \frac{1}{2}
\begin{pmatrix}
1 \pm e^{-ik} \\
1 \mp e^{-ik}
\end{pmatrix}\,.
\end{equation}
The Bloch functions are polynomials in $e^{\pm ik}$, therefore their Fourier transform~(\ref{eq:Wannier_func_def}) produces compact Wannier functions
\begin{equation}\label{eq:Wannierf_Creutz}
w_{\pm}(j) = \begin{cases}
\frac{1}{2}(1,1)^T & j = 0\,, \\
\pm \frac{1}{2}(1,-1)^T & j = 1\,, \\
(0,0)^T & j \neq 0,1\,.
\end{cases}
\end{equation}

\subsection{Dice lattice}
\label{app:bands_dice_lattice}

The band structure of the dice lattice with hopping matrix elements given by Eq.~(\ref{eq:2D_dice_def}) is composed of three doubly degenerate flat bands with energy eigenvalues 
\begin{gather}
\varepsilon_0 = 0\,,\label{eq:energy_dice1}\\
\varepsilon_{\pm} = \frac{1}{2}\left(\varepsilon_{\rm h} \pm \sqrt{\varepsilon_{\rm h}^2+4|t_1|^2+8|t_2|^2+8|t_3|^2+4|t_4|^2}
\right)\,.\label{eq:energy_dice2}
\end{gather}
In the following we label the bands in increasing energy order. The lower bands with energy $\varepsilon_-$ are labelled by $n = 1,2$, the middle bands with energy $\varepsilon_0$ by $n = 3,4$ and the upper ones with energy $\varepsilon_{+}$ by $n = 5,6$. The polynomial Bloch functions for the upper and lowest bands have a simple form for arbitrary values of the parameters $t_{i = 1,\dots,4}$ and $\varepsilon_{\rm h}$ and are given by
\begin{gather}
g_1(\vc{k}) = c(t_3^*+t_2e^{-ik_1},t_2^*-t_3e^{-ik_1}, \varepsilon_-, 0 , e^{ik_2}t_4^*,t_1)^T\,,\label{eq:g_1_dice}\\
g_2(\vc{k}) = c(-t_1^*,t_4e^{-ik_2},0,\varepsilon_-,-t_2+t_3^*e^{ik_1},t_3+t^*_2e^{i k_1})^T\,.\label{eq:g_2_dice}
\end{gather}
The normalization factor $c$ is given by
\begin{equation}\label{eq:norm_fact}
c = \frac{1}{\sqrt{\varepsilon_{-}^2 + |t_1|^2+2|t_2|^2+2|t_3|^2+|t_4|^2}}\,.
\end{equation}
Since the normalization factor does not depend on $\vc{k}$ the Fourier transform of Eq.~(\ref{eq:g_1_dice})-(\ref{eq:g_2_dice}) produces a complete orthonormal set of Wannier functions for the doubly degenerate lower flat bands. 
The corresponding Bloch functions for the upper bands are obtained from those of the lowest band with the substitution $\varepsilon_-\to \varepsilon_+$ in Eqs.~(\ref{eq:g_1_dice})-(\ref{eq:norm_fact}). We choose the convention that under this substitution the Bloch functions are mapped as $g_1(\vc{k})\to g_5(\vc{k})$ and $g_2(\vc{k})\to g_6(\vc{k})$. The compact Wannier functions obtained from $g_{1,2}(\vc{k})$ and $g_{5,6}(\vc{k})$ have support in the smallest number of lattice sites (seven lattice sites, one hub site together with the neighbouring rim sites).

For the middle bands we are not able to provide an analogous orthonormal basis composed of compact Wannier function in two dimensions. We can at best provide Wannier functions that are orthogonal and compact along only a chosen spatial direction, but only exponentially decaying in the orthogonal direction. Although it is not difficult to obtain them for arbitrary values of the parameters $t_i$ in Eq.~(\ref{eq:2D_dice_def}) we only need the special case $t_{i = 1,2,3,4}=t$ to present the conserved quantities of the 1D dice lattice as parity operators in the Wannier function basis (Eq.~(\ref{eq:Cj_Def_dice}) for $k_2 =0$ and Eq.~(\ref{eq:parity_1}) with $n =0,\dots,6$ for $k_2 =\pi$) . Therefore we provide them only in this latter case. The Bloch functions that produce the required Wannier functions are then
\begin{gather}
g_3(\vc{k}) = \begin{pmatrix}
6 \left(-1+e^{-i k_1}\right)-\left(1+e^{-i k_1}\right) \left(e^{-i k_2}+e^{i k_2}\right)\\
-6 \left(1+e^{-i k_1}\right) + \left(-1+e^{-i k_1}\right) \left(e^{-i k_2}+e^{i k_2}\right)\\0\\0\\5-e^{2 i k_2}\\
5 e^{i k_2}-e^{-i k_2}
\end{pmatrix}\\
g_4(\vc{k}) = \begin{pmatrix}
5 e^{-i k_2}-e^{i k_2}\\
-5+e^{-2 i k_2}\\0\\0\\-6 \left(1+e^{i k_1}\right) + \left(-1+e^{i k_1}\right) \left(e^{-i k_2}+e^{i k_2}\right)\\
6 \left(1-e^{i k_1}\right)+\left(1+e^{i k_1}\right) \left(e^{-i k_2}+e^{i k_2}\right)
\end{pmatrix}\,.
\end{gather}
These Bloch function are not normalized since
\begin{equation}\label{eq:dice_norm}
|g_3(\vc{k})|^2 = |g_4(\vc{k})|^2 = 6(6+\cos k_2)(6-\cos k_2)\,.
\end{equation}
Therefore the corresponding normalized Bloch functions are polynomials in $e^{\pm ik_1}$, but not in $e^{\pm ik_2}$, and their Fourier transform produces compact Wannier functions only along direction $\vc{a}_1$ of the Bravais lattice, but not along $\vc{a}_2$ where they are just exponentially decaying. If the Fourier transform of the unnormalized Bloch functions is taken, one obtains compact functions whose translation along $\vc{a}_2$ are not orthogonal. This in a example of the more general notion of compactly supported Wannier-type functions introduced in Ref.~\cite{Read:2017}. If the Fourier transform is taken only with respect to $k_1$, one obtains a basis of orthogonal and compact Wannier functions which are eigenstates of the 1D dice lattice of Fig.~\ref{fig:lattice_models}\textbf{c} for arbitrary values of $k_2$. These compact Wannier functions are used to present the conserved quantities of the 1D dice lattice as parity operators in the Wannier function basis (Eq.~(\ref{eq:Cj_Def_dice}) for $k_2 =0$ and Eq.~(\ref{eq:parity_1}) with $n =0,\dots,6$ for $k_2 =\pi$) and they are in some sense the best possible choice since they have the smallest possible support in terms of number of lattice sites and they are uniquely specified (up to normalization) given the support. Similar Bloch functions that are polynomials only in $e^{\pm ik_2}$ but not in $e^{\pm ik_1}$ can be found but are not presented in here.

\subsection{Diamond chain}
\label{app:bands_diamond_chain}

The energy of the three flat bands of the diamond chain are obtained by setting $t_1 = t_4 = 0$ in Eq.~(\ref{eq:energy_dice1})-(\ref{eq:energy_dice2}). The corresponding polynomial Bloch functions are then
\begin{gather}
g_\pm(k) = c_\pm(t_3^*+t_2e^{-ik a}, t_2^*-t_3e^{-ik a},\varepsilon_\pm)^T\,,\\
c_\pm = \frac{1}{\sqrt{\varepsilon_\pm^2+2|t_2|^2+2|t_3|^2}}\,,\\
g_0(k) = (t_3^*-t_2e^{-ik a}, t_2^*+t_3e^{-ik a},0)^T\,.
\end{gather}
Their Fourier transform produces a complete set of orthornormal compact Wannier functions for the diamond chain for arbitrary values of the parameters $t_2,t_3,\varepsilon_{\rm h}$. In Fig.~\ref{fig:lattice_models}\textbf{d} the special case $t_1 = t_2 = t$ is shown. In the main text we refer only to this special case when results regarding the diamond lattice are presented.

\section{Global intertwining operators}
\label{app:global_U}

Here we present the global intertwining operators for the Creutz ladder, the diamond chain and the 1D and 2D dice lattices. These have been mentioned in Sec.~\ref{sec:graph_automorphism}. In the case of the Creutz ladder the noninteracting Hamiltonian is given by
\begin{equation}
\mathcal{\hat H}_0 = \sum_{j=0}^{N_{\rm c}-1}\hat{\vc{c}}_{j+1}^\dagger K(1) \hat{\vc{c}}_{j} +\text{H.c.}\,,\quad K(1) = t_1\sigma_z + it_2 \sigma_y\,. 
\end{equation}
Consider the following canonical transformation
\begin{equation}\label{eq:Creutz_global}
\hat{\vc{c}}_{j} \to (-1)^j\sigma_x \hat{\vc{c}}_{j}\,.
\end{equation}
For $j \neq N_{\rm c}-1$ we have
\begin{equation}
\hat{\vc{c}}_{j+1}^\dagger K(1) \hat{\vc{c}}_{j} \to (-1)^{2j+1}
\hat{\vc{c}}_{j+1}^\dagger \sigma_x K(1) \sigma_x\hat{\vc{c}}_{j}
= \hat{\vc{c}}_{j+1}^\dagger K(1) \hat{\vc{c}}_{j}\,,
\end{equation}
while for $j = N_{\rm c}-1$
\begin{equation}
\hat{\vc{c}}_{0}^\dagger K(1) \hat{\vc{c}}_{N_{\rm c}-1} \to (-1)^{N_{\rm c}}\hat{\vc{c}}_{0}^\dagger K(1) \hat{\vc{c}}_{N_{\rm c}-1}\,.
\end{equation}
Thus for a chain of even length the above transformation leaves the Hamiltonian invariant, while in the odd length case one obtains an additional minus sign on the term $\hat{\vc{c}}_{0}^\dagger K(1) \hat{\vc{c}}_{N_{\rm c}-1}$, that is the Hamiltonian with flux $\Phi = 0$ is mapped into the one with flux $\Phi = \pi$. The transformation~(\ref{eq:Creutz_global}) involves all lattice sites and corresponds geometrically to a reflection that exchanges the orbitals $\alpha = 1\leftrightarrow \alpha = 2$ with each unit cell. The Hubbard interaction term (and even more general types of interactions) is left invariant by this transformation. Therefore Eq.~(\ref{eq:Creutz_global}) defines an intertwining operator for a Creutz ladder of odd length. 
The intertwining operator for the diamond lattice is very similar and is constructed from a reflection which swaps orbitals $\alpha =1$ and $\alpha = 2$ within all unit cells (see below).
 
In the case of the 1D and 2D dice lattices one considers canonical transformations defined in terms of the following matrix
\begin{equation}
A(k_2) = \begin{pmatrix}
0 & e^{ik_2} & 0 & 0 & 0 & 0 \\
e^{ik_2} & 0 & 0 & 0 & 0 & 0 \\
0 & 0 & e^{ik_2} & 0 & 0 & 0 \\
0 & 0 & 0 & -1 & 0 & 0\\
0 & 0 & 0 & 0 & 0 & 1\\
0 & 0 & 0 & 0 & 1 & 0
\end{pmatrix}\,.
\end{equation} 
One can easily show that, for $t_1 = t_2 = t_3 = t_4 = t$ and $t$ real in Eq.~(\ref{eq:2D_dice_def}), 
\begin{equation}\label{eq:A_Tk}
A^\dagger(k_2) \widetilde{K}(k_1,k_2)A(k_2) = \widetilde{K}(k_1+\pi,-k_2)\,.
\end{equation}
In the case of the 1D dice lattice, Eq.~(\ref{eq:A_Tk}) implies that the action of $A(k_2)$ on $K_{k_2}(j)$ defined by Eq.~(\ref{eq:partial_FT}) is 
\begin{equation}
A^\dagger(k_2) K_{k_2}(j)A(k_2) = (-1)^jK_{-k_2}(j)\,.
\end{equation}
Let $\mathcal{\hat U}_{\rm g}$ the unitary operator which transforms the field operators of the 1D dice lattice as $\mathcal{\hat U}_{\rm g} \hat{\vc{c}}_j \mathcal{\hat U}_{\rm g}^\dagger = A(k_2) \hat{\vc{c}}_j$. 
Then its action on the noninteracting Hamiltonian is
\begin{equation}
\begin{split}
&\mathcal{\hat U}_{\rm g}\mathcal{\hat H}_{0,k_2}(\Phi = 0) \mathcal{\hat U}_{\rm g}^\dagger = \sum_{i,j = 0}^{N_{\rm c}-1} \mathcal{\hat U}_{\rm g}\hat{\vc{c}}^\dagger_i K_{k_2}(i-j) \hat{\vc{c}}_j \mathcal{\hat U}_{\rm g}^\dagger \\
&= \sum_{j = 0}^{N_{\rm c}-1} (-1)^{i-j}\hat{\vc{c}}^\dagger_i K_{-k_2}(i-j)\hat{\vc{c}}_j = \mathcal{\hat H}_{0,-k_2}(\Phi = N_{\rm c} \pi)\,.
\end{split}
\end{equation}
The factor $(-1)^{i-j}$ in the second line in the above equation introduces a total magnetic flux $\Phi = N_{\rm c} \pi$ threading the ring. This is equivalent to $\Phi = 0 $ for even length and $\Phi = \pi$ for odd length. Note that we have mapped 
the noninteracting Hamiltonian $\mathcal{\hat H}_{0,k_2}$ to its time-reversal partner with $k_2\to -k_2$. In order to recover the original Hamiltonian with an additional $\Phi = \pi$ flux one has to use the antiunitary time-reversal operator $\mathcal{\hat T}$
\begin{equation}
\mathcal{\hat T}\mathcal{\hat H}_{0,-k_2}\mathcal{\hat T}^{-1} = \mathcal{\hat H}_{0,k_2}\,.
\end{equation}
The complete global intertwining operator is thus $\mathcal{\hat T}\mathcal{\hat U}_{\rm g}$, modulo a gauge transformation. It is again evident that this operator preserves the Hubbard term and other more general types of interaction terms and is therefore an intertwining operator for the full many-body Hamiltonian.

The intertwining operator for the 2D dice lattice is closely related to the one of the 1D dice lattice. The only subtle point is what is the effect of the time-reversal operator once the Fourier transform with respect to $k_2$ is performed as well. One expects that there would be no need for the time-reversal operator since the Hamiltonian of the 2D dice lattice in Fig.~\ref{fig:lattice_models}\textbf{b} is time-reversal invariant. Indeed, the answer is that after Fourier transforming the time-reversal operator becomes a spatial reflection operator. It is easier to provide directly the global intertwining operator for the 2D dice lattice in real space. It reads
\begin{gather}
\mathcal{\hat U}_{\rm g} \hat{c}_{(i_1,i_2),1}\mathcal{\hat U}_{\rm g}^\dagger = \hat{c}_{(i_1,1-i_2),2}\,,\label{eq:diamond_1}\\ \mathcal{\hat U}_{\rm g} \hat{c}_{(i_1,i_2),2}\mathcal{\hat U}_{\rm g}^\dagger = \hat{c}_{(i_1,1-i_2),1}\,,\label{eq:diamond_2}
\end{gather}
\begin{gather}
\mathcal{\hat U}_{\rm g} \hat{c}_{(i_1,i_2),3}\mathcal{\hat U}_{\rm g}^\dagger = \hat{c}_{(i_1,1-i_2),3}\,,\label{eq:diamond_3}\\
\mathcal{\hat U}_{\rm g} \hat{c}_{(i_1,i_2),4}\mathcal{\hat U}_{\rm g}^\dagger = -\hat{c}_{(i_1,-i_2),4}\,,\\
\mathcal{\hat U}_{\rm g} \hat{c}_{(i_1,i_2),5}\mathcal{\hat U}_{\rm g}^\dagger = \hat{c}_{(i_1,-i_2),6}\,,\\ \mathcal{\hat U}_{\rm g} \hat{c}_{(i_1,i_2),6}\mathcal{\hat U}_{\rm g}^\dagger = \hat{c}_{(i_1,-i_2),5}\,.
\end{gather}
One can easily check that this intertwining operator is built out of a spatial reflection ($i_2\to -i_2$), which is an element of the space group of the dice lattice. For the 2D dice lattice the graph autorphism group and the space group coincide, as mentioned in Sec.~\ref{sec:graph_automorphism}. If one ignores the index $i_2$, Eqs.~(\ref{eq:diamond_1})-(\ref{eq:diamond_3}) give the global intertwining operator of the diamond chain, for $t_2 = t_3 = t$ real in Eq.~(\ref{eq:diamond_T_def}).

%

\end{document}